\documentclass[prb,twocolumn,longbibliography,superscriptaddress,preprintnumbers]{revtex4-2}
\usepackage{bm}
\usepackage{dsfont}
\usepackage{graphicx}
\usepackage{epsfig}
\usepackage{epstopdf}
\usepackage{balance}
\usepackage[dvipsnames]{xcolor}
\usepackage{calc}
\usepackage{natbib}
\usepackage[colorlinks,
            linkcolor=blue,
            anchorcolor=blue,
            citecolor=blue,
            urlcolor=blue]{hyperref}
\usepackage{lipsum}
\usepackage[version=3]{mhchem} 
\usepackage{color}
\usepackage{soul}
\usepackage[etex=true,export]{adjustbox}
\usepackage{makecell}
\usepackage{multirow}
\usepackage{tabularx,multirow,array,diagbox}
\usepackage{adjustbox}
\usepackage{booktabs}
\makeatletter
\renewcommand{\maketag@@@}[1]
{\hbox{\m@th\normalsize\normalfont#1}}%
\newcommand\red[1]{{#1}}
\makeatother
  
\begin{document}
\title{Variational Mapping of Chern Bands to Landau Levels: Application to Fractional Chern Insulators in Twisted MoTe$_2$}

\author{Bohao Li}
\affiliation{School of Physics and Technology, Wuhan University, Wuhan 430072, China}
\author{Fengcheng Wu}
\email{wufcheng@whu.edu.cn}
\affiliation{School of Physics and Technology, Wuhan University, Wuhan 430072, China}
\affiliation{Wuhan Institute of Quantum Technology, Wuhan 430206, China}

\begin{abstract}
We present a theoretical study of mapping between Chern bands and generalized Landau levels in twisted bilayer MoTe$_2$ \red{($t$MoTe$_2$)}, where fractional Chern insulators down to zero magnetic fields have been observed. We construct an exact Landau-level representation of moir\'e bands, where the basis functions, characterized by a uniform quantum geometry, are derived from Landau-level wavefunctions dressed by spinors aligned or antialigned with the layer pseudospin skyrmion field. We further generalize the dressed zeroth Landau level to a variational wavefunction with an ideal yet nonuniform quantum geometry and variationally maximize its weight in the first moir\'e band. The variational wavefunction has a high overlap with the first band and quantitatively captures the exact diagonalization spectra of fractional Chern insulators at hole-filling factors $\nu_h=2/3$ and $3/5$, providing a clear theoretical mechanism for the formation and properties of the fractionalized states. Our work introduces a variational approach to studying fractional states by mapping Chern bands to Landau levels, \red{with application to other systems beyond $t$MoTe$_2$ also demonstrated.} 
\end{abstract}
\maketitle

\section{introduction}
Fractional Chern insulators (FCIs), proposed in theory \cite{Tang2011,Sun2011,Neupert2011,Regnault2011Fractional,Sheng2011Fractional} as lattice generalizations of fractional quantum Hall states in Landau levels (LLs), have now been observed in several van der Waals heterostructures \cite{Spanton2018Observation,Xie2021Fractional,Cai2023,Zeng2023,Park2023,Xu2023Observation,lu2024fractional}. Among them, FCIs at zero magnetic field, also known as fractional quantum anomalous Hall insulators, were first realized in $t$MoTe$_2$ \cite{Cai2023,Zeng2023,Park2023,Xu2023Observation}. Transport studies of $t$MoTe$_2$ have identified both integer Chern insulators at $\nu_h=1$ and FCIs at $\nu_h=2/3$ and $3/5$ \cite{Park2023,Xu2023Observation}, further characterized through the optical spectrum \cite{Cai2023,Zeng2023}, electronic compressibility \cite{Zeng2023} and real-space local imaging \cite{ji2024local,redekop2024direct}. In addition, a recent experiment has reported evidence of fractional quantum spin Hall effect in $t$MoTe$_2$ at $\nu_h=3$ \cite{kang2024evidence}.
While earlier theoretical studies predicted Chern bands \cite{Wu2019Topological,Yu2019Giant,Devakul2021} and FCIs \cite{Li2021Spontaneous,Crepel2023,Nicolas2023Pressure} in twisted transition metal dichalcogenide homobilayers, the experimental breakthroughs have sparked many further theoretical investigations into the rich quantum phases of matter in this system \cite{Goldman2023Zero,Dong2023,Reddy2023Toward,Reddy2023Fractional,Wang2024Fractional,Xu2024Maximally,mao2023lattice,Zhang2024Polarization,Jia2024Moire,Qiu2023,Li2024Electrically,Luo2024,yu2024fractional,Abouelkomsan2024,Song2024,Fan2024Orbital,Liu2024Gate,Nicolas2024Magic,shi2024adiabatic,Crepel2024Chiral,wang2023topology,reddy2024nonabelian,xu2024multiple,ahn2024landau,wang2024higher,zhang2024nonabelian,jian2024minimal,villadiego2024halperin,maymann2024theory,wang2024interacting,Sankar2024zero}.

The microscopic mechanism underlying the formation of FCIs is a fundamental theoretical question. One approach involves establishing connections between Bloch Chern bands and LLs \cite{Qi2011Generic,Wu2012Gauge,Siddharth2013,Jackson2015,Claassen2015,Tarnopolsky2019,Ledwith2020,Wang2021Chiral,Wang2021Exact,Ozawa2021Relations,Mera2021Kahler,Wang2022Hierarchy,Ledwith2022IdealChern,Ledwith2023Vortexability,Wang2023Origin,Dong2023Manybody,fujimoto2024higher}. In $t$MoTe$_2$, the first moir\'e valence band has a Chern number $|\mathcal{C}|=1$ and nearly, though not exactly, saturates the trace inequality for quantum geometry at a magic twist angle $\theta_m$, signaling similarities with the zeroth LL (0LL) \cite{Dong2023}. Remarkably, the exact-diagonalization (ED) spectra of $t$MoTe$_2$, with the many-body Hamiltonian projected onto the first band, resembles those of 0LL, aiding in the numerical determination of FCIs at $\nu_h$ with odd denominators and composite Fermi liquid at $\nu_h=1/2$ \cite{Goldman2023Zero,Dong2023,Reddy2023Fractional,Reddy2023Toward}. Connections between topological moir\'e bands in $t$MoTe$_2$ and LLs have been made in different approximate limits \cite{Nicolas2024Magic,shi2024adiabatic,Crepel2024Chiral}, but a quantitative mapping strategy in the generic case is still lacking.

Here we introduce a strategy that begins with an exact LL representation of moir\'e bands in $t$MoTe$_2$. The construction is motivated by the layer pseudospin field, which forms a skyrmion lattice \cite{Wu2019Topological} and generates a non-Abelian gauge field in the local frame \cite{Yu2019Giant,Nicolas2024Magic}. The Bloch bases in this representation consist of  \textit{dressed} LL wavefunctions [Eq.~\eqref{phidecom}] that maintain uniform quantum geometry. The average weight of the first moir\'e band on the dressed 0LL states is considerable ($\sim 0.75$) but noticeably below unity at $\theta_m$. We then extend the dressed 0LL states to generalized 0LL wavefunctions [Eq.~\eqref{wavef1}] with an ideal (i.e., saturated trace inequality) but nonuniform quantum geometry \cite{Wang2021Exact} and optimize their weight in the first moir\'e band using a variational approach, reaching a maximum of $0.95$ at $\theta_m$. The generalized 0LL wavefunctions capture the density fluctuation in the Chern band and host exact FCI ground state for certain short-range interactions. The obtained high weight provides an intuitive explanation of the emergence of FCIs in $t$MoTe$_2$.   We also perform ED studies with Coulomb interactions projected onto both the original wavefunction of the first band and the variational wavefunction. Both models consistently yield FCIs at $\nu_h=2/3$ and $3/5$, with a quantitative agreement in the energy spectra. Based on the variational wavefunction, we quantitatively examine the effect of density fluctuation on FCIs, explaining their twist angle dependence. \red{We further demonstrate the mapping in a model system \cite{tan2024designing} beyond $t$MoTe$_2$, showing the generic applicability of our approach. }This mapping strategy enables the application of physics from fractional quantum Hall states to FCIs and the design of new systems that host FCIs.

\red{This paper is organized as follows. In Sec. \ref{section:2}, we present the single-particle Hamiltonian and band structure of $t$MoTe$_2$. In Sec. \ref{section:3}, we construct the dressed LL bases for the first moir\'e Chern band. Section \ref{section:4} develops the variational approach for mapping the Chern band to the generalized 0LL. In Sec. \ref{section:5}, we perform ED studies of FCIs, using both the original and variational wavefunctions. Section \ref{section:6} explores further applications in a model system without the emergent magnetic field. Finally, we conclude with a discussion in Sec. \ref{section:7}. Technical details are provided in Appendices \ref{appendix:A} and \ref{appendix:B}.}

\section{Moir\'e Hamiltonian}
\label{section:2}
Moir\'e Hamiltonian for valence states in \textit{t}MoTe$_2$ at $+K$ valley is given by \cite{Wu2019Topological,Jia2024Moire},
\begin{equation}
\begin{aligned} 
H = & \begin{pmatrix}
-\frac{\hbar^2(\hat{\boldsymbol k}-\boldsymbol {\kappa}_+)^2}{2m^*}+\Delta_+(\boldsymbol r) &   \Delta_{t}(\boldsymbol r)
\\ \Delta_{t}^\dagger(\boldsymbol r)
 & -\frac{\hbar^2(\hat{\boldsymbol k}-\boldsymbol{\kappa}_-)^2}{2m^*}+\Delta_-(\boldsymbol r)
\end{pmatrix},
\end{aligned}
\end{equation}
where $\hat{\boldsymbol k}$ is the momentum operator, $m^*$ is the effective mass, $\boldsymbol{\kappa}_\pm=\frac{4\pi}{3a_M}(-\frac{\sqrt{3}}{2},\mp\frac{1}{2})$ are located at corners of moir\'e Brillouin zone, $a_M \approx a_0/\theta$ is the moir\'e period, $\theta$ is the twist angle, and $a_0=3.52 ~\text{\AA}$ is the monolayer lattice constant. $\Delta_{\pm}(\boldsymbol{r})$ and $\Delta_t(\boldsymbol{r})$ are given by
\begin{equation} 
\begin{aligned}
\Delta_\pm(\boldsymbol r)=& 2V_{1}\!\!\sum_{j=1,3,5}\!\!\cos(\boldsymbol{g}^{(1)}_{j}\!\cdot\! \boldsymbol{r}\!\pm\!\psi)+2V_{2}\!\!\sum_{j=1,3,5}\!\!\cos(\boldsymbol{g}^{(2)}_{j}\!\cdot\! \boldsymbol{r}),\\
\Delta_{t}(\boldsymbol r)= & w_1(1+e^{-i\boldsymbol{g}^{(1)}_{2}\cdot\boldsymbol r}+e^{-i\boldsymbol{g}^{(1)}_{3}\cdot\boldsymbol r}) \\
 & + w_2(e^{-i\boldsymbol{g}^{(2)}_2\cdot \boldsymbol{r}}+e^{-i\boldsymbol{g}^{(1)}_1\cdot \boldsymbol{r}}+e^{-i\boldsymbol{g}^{(1)}_4\cdot \boldsymbol{r}}),
\end{aligned}
\end{equation}    
where  $\boldsymbol{g}_{i}^{(1)}=\frac{4\pi}{\sqrt{3}a_M}[\cos\frac{\pi(i-1)}{3},\sin\frac{\pi(i-1)}{3}]$ and $\boldsymbol{g}_{i}^{(2)}= \frac{4\pi}{a_M}[\cos\frac{\pi(2i-1)}{6},\sin\frac{\pi(2i-1)}{6}] $ are moir\'e reciprocal lattice vectors. We use  model parameters fitted from first-principles band structure  at $\theta=3.89^\circ$~\cite{Jia2024Moire}, where the effective mass $m^*=0.62\,m_e$ ($m_e$ is the electron rest mass) and $(\psi,V_1,w_1,V_2,w_2)=(-88.43^\circ,7.94\,\rm{meV},-10.77\,\rm{meV},20.00\,\rm{meV},10.21\,\rm{meV})$. 
Here we use this particular set of parameters to illustrate the main physics \cite{Reddy2023Fractional,Wang2024Fractional,Xu2024Maximally,mao2023lattice,Zhang2024Polarization,Jia2024Moire}.

\begin{figure}[t]
    \includegraphics[width=1.\columnwidth]{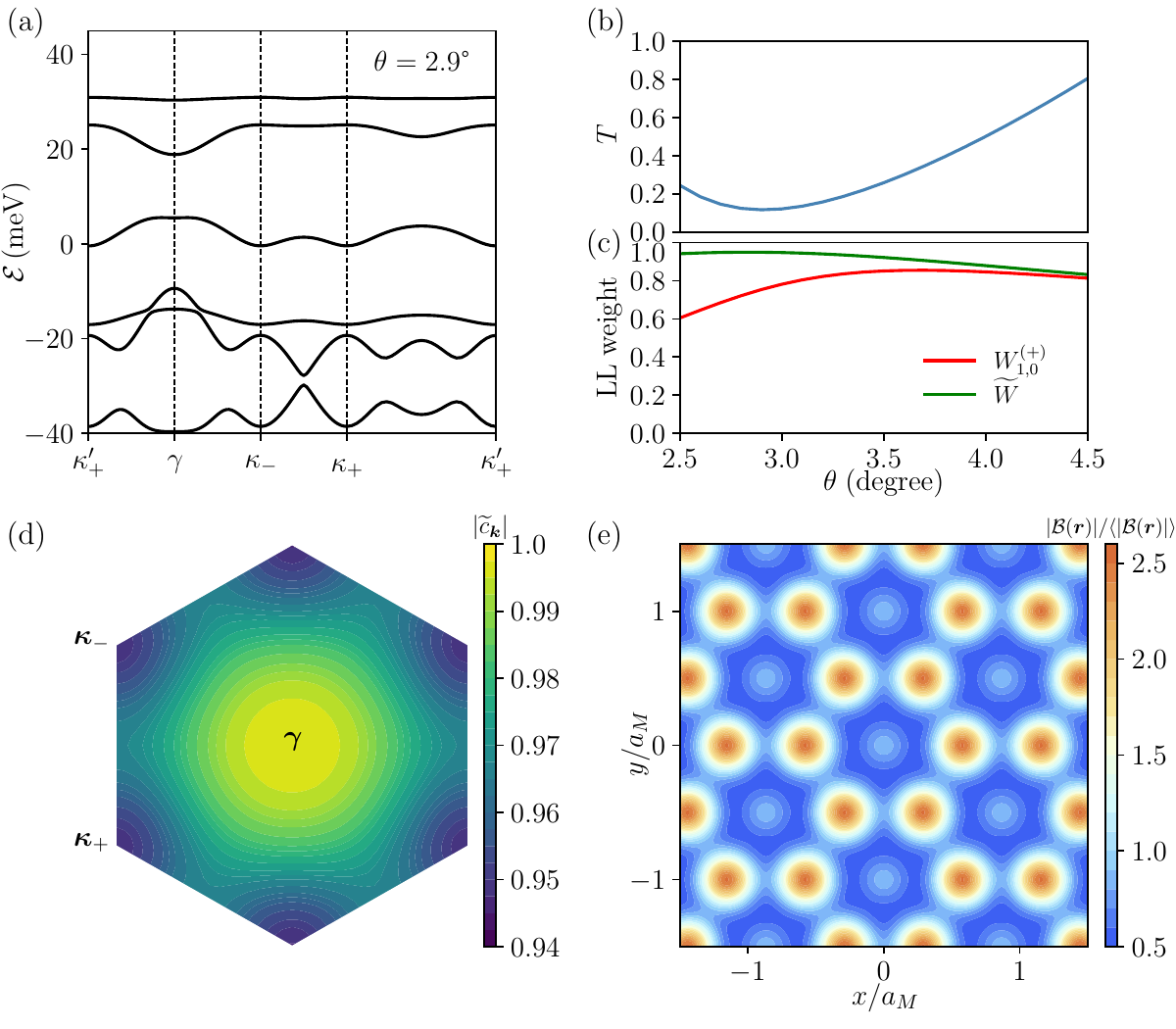}
    \caption{(a) Moir\'e band structure of \textit{t}MoTe$_2$ at $\theta=2.9^\circ$. (b) $T$ of the first band as a function of $\theta$. (c) LL weight $W_{1,0}^{(+)}$ (red line) and $\widetilde W$ (green line) as functions of $\theta$.  (d) Overlap $\lvert \widetilde c_{\boldsymbol{k}}\rvert = \lvert \langle\varphi_{1,\boldsymbol k}\lvert\Theta_{\boldsymbol k}\rangle \rvert$  in moir\'e Brillouin zone. (e) Map of $\lvert \mathcal{B}(\boldsymbol{r})\rvert$ scaled by its spatial average. $\theta$ is $2.9^\circ$ in (d) and (e).} 
    \label{fig:1}
\end{figure}

\begin{figure}[t]    \centerline{\includegraphics[width=1.05\columnwidth]{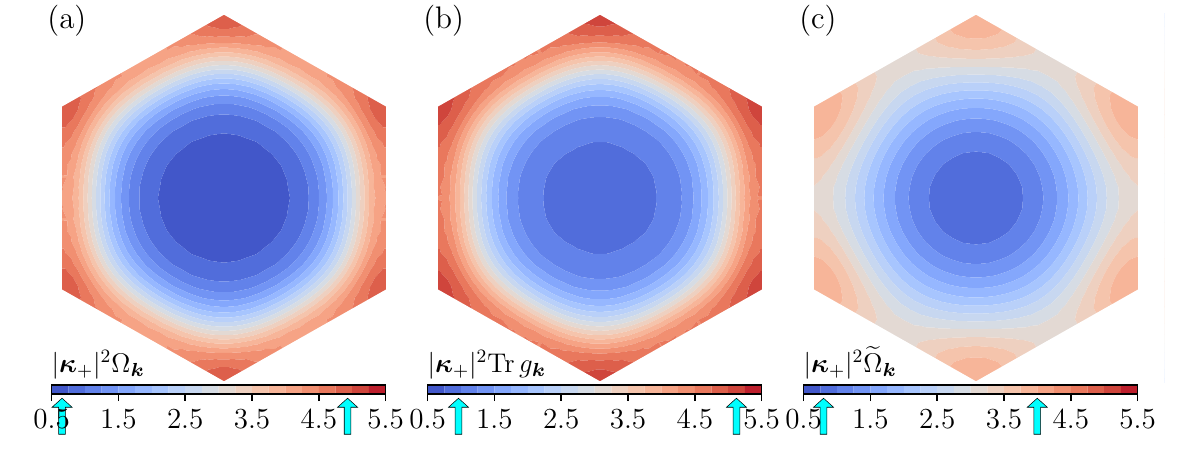}}
    \caption{(a) $\Omega_{\boldsymbol{k}}$ and (b) $\text{Tr}\,g_{\boldsymbol{k}}$ for  $\varphi_{1,\boldsymbol  k}(\boldsymbol r)$ in the moir\'e Brillouin zone. (c) $\widetilde\Omega_{\boldsymbol{k}}$ (identical to $\text{Tr}\,\widetilde g_{\boldsymbol{k}}$) for $\Theta_{\boldsymbol  k}(\boldsymbol r)$. The arrows indicate the data range in each plot. $\theta$ is $2.9^{\circ}$. }
    \label{fig:2}
\end{figure}

A representative moir\'e band structure is shown in Fig.~\ref{fig:1}(a). We focus on the first moir\'e valence band with a narrow bandwidth and a Chern number $\mathcal {C}=+1$. The quantum geometry of this band, as characterized by Berry curvature $\Omega_{\boldsymbol{k}}$ and trace of quantum metric $\text{Tr}\,g_{\boldsymbol{k}}$, is illustrated in Fig.~\ref{fig:2} for $\theta=2.9^{\circ}$. Here $\Omega_{\boldsymbol{k}}$ is positive definite and fluctuates in sync with $\text{Tr}\,g_{\boldsymbol{k}}$. A measure of the deviation from ideal quantum geometry \cite{Ledwith2020,Wang2021Exact} is $T=\frac{1}{2\pi}\int d^2\boldsymbol k\mathrm{Tr}\,g_{\boldsymbol{k}}-|\mathcal{C}|$, which is bounded by $T \ge 0$ (i.e., the trace inequality). The $\theta$ dependence of $T$ is plotted in Fig.~\ref{fig:1}(b), which is minimum $(T\sim 0.1)$  at the magic angle $\theta_m=2.9^{\circ}$. The small value of $T$ at $\theta_m$ implies a connection with the generalized 0LL, which has a nonuniform but ideal quantum geometry (i.e., $T=0$).

\section{Landau Level representation}
\label{section:3}
\begin{figure}[t]
    \includegraphics[width=1.\columnwidth]{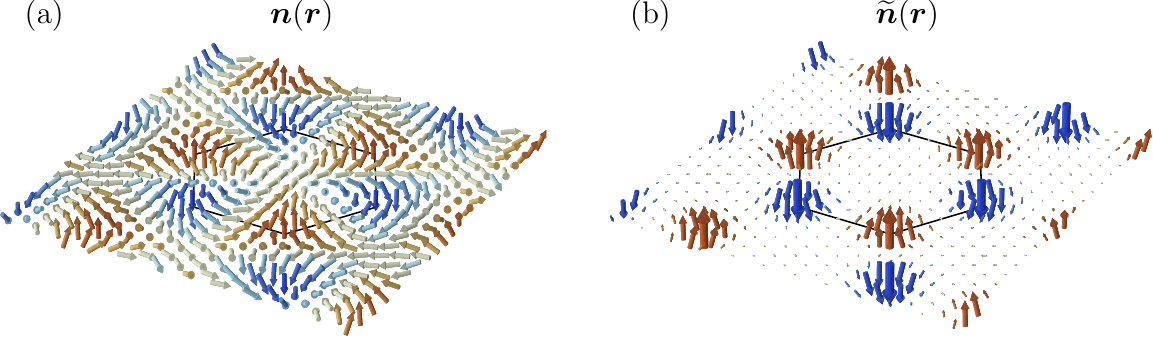}
    \caption{(a), (b) Plot of $\boldsymbol n(\boldsymbol{r})$ and $\widetilde{\boldsymbol{n}}(\boldsymbol{r})$, with color encoding the $\hat{z}$ component. The hexagon marks the Wigner-Seitz cell.}
    \label{fig:3}
\end{figure}

We apply two unitary transformations to $H$ to find a representation of Bloch states in terms of LL wavefunctions. 
The first unitary transformation is to recast the Hamiltonian into the form of a particle moving in a scalar potential $\Delta_0(\boldsymbol r)$ and a layer pseudospin field $\boldsymbol\Delta(\boldsymbol r)$  that forms a skyrmion lattice,
\begin{equation}
\begin{aligned}
  H_1 &=  U_{0}^\dagger(\boldsymbol r)HU_{0}(\boldsymbol r),  \\
 &= -\frac{\hbar^2\hat{\boldsymbol k}^2}{2m^*}\sigma_0+\boldsymbol \Delta(\boldsymbol r)\cdot \boldsymbol\sigma+\Delta_0(\boldsymbol r)\,\sigma_0, 
\label{A1}
\end{aligned}
\end{equation}
where $U_0(\bm r)=\text{diag}(e^{i\boldsymbol \kappa_+\cdot \boldsymbol r},e^{i\boldsymbol \kappa_-\cdot \boldsymbol r})$, $\sigma_0$ is the identity matrix and $\boldsymbol{\sigma}$ are Pauli matrices. 
The skyrmion field  $\boldsymbol\Delta(\boldsymbol r)$ has a nontrivial winding number,
\begin{align}
N_w=&\frac{1}{2\pi}\int_{\mathcal{A}_0}d\boldsymbol{r}\,b_z =-1. 
\label{Nw}
\end{align}
Here $\mathcal{A}_0$ is the area expanded by a moir\'e unit cell, $b_z=\frac{1}{2}\boldsymbol n\cdot(\partial_x\boldsymbol n\times\partial_y\boldsymbol n)$, and  $\boldsymbol{n}(\boldsymbol{r})=\boldsymbol\Delta(\boldsymbol r)/\lvert\boldsymbol\Delta(\boldsymbol r)\rvert$. An illustration of the skyrmion lattice formed by the unit vector $\boldsymbol{n}(\boldsymbol{r})$ is shown in Fig.~\ref{fig:3}(a).

The skyrmion field generates an emergent magnetic field in the adiabatic limit \cite{Nagaosa2013}, which facilitates the mapping to LLs.
We construct another unitary matrix $U(\boldsymbol r)$ to rotate the layer pseudospin to the local frame of $\boldsymbol \Delta(\boldsymbol r)$,
\begin{equation}
\begin{aligned}
&U^\dagger(\boldsymbol r)[\boldsymbol \Delta(\boldsymbol r)\cdot \boldsymbol \sigma] U(\boldsymbol r)= \lvert\boldsymbol\Delta(\boldsymbol r)\rvert\sigma_z,\\
&U(\boldsymbol r)= (\boldsymbol\chi^{(+)}(\boldsymbol r),\boldsymbol\chi^{(-)}(\boldsymbol r)),
\end{aligned}
\end{equation}
where $\boldsymbol\chi^{(+)}$ and $\boldsymbol\chi^{(-)}$ are spinors that align and antialign with $\boldsymbol \Delta$, respectively.
The nonzero winding number $N_w$ poses an obstruction in choosing the phases of $\boldsymbol\chi^{(\pm)}$ such that $U(\boldsymbol r)$ is spatially continuous and periodic. However, $U(\boldsymbol r)$ can be continuous but quasiperiodic. This motivates us to express $U(\boldsymbol r)$ using LL wavefunctions,
\begin{equation}
\label{chi_mag}
\begin{aligned}
\boldsymbol\chi^{(+)}(\boldsymbol r)=&
\{\alpha(\boldsymbol r)\Psi^{(+)}_{0,\boldsymbol \kappa_-}(\boldsymbol r),
\beta(\boldsymbol r)\Psi^{(+)}_{0,\boldsymbol \kappa_+}(\boldsymbol r)\}^T,\\
\boldsymbol\chi^{(-)}(\boldsymbol r)=&
\{\beta^*(\boldsymbol r)[\Psi^{(+)}_{0,\boldsymbol \kappa_+}(\boldsymbol r)]^*,
-\alpha^*(\boldsymbol r)[\Psi^{(+)}_{0,\boldsymbol \kappa_-}(\boldsymbol r)]^*\}^T.
\end{aligned}
\end{equation}
Here $\Psi_{n,\boldsymbol  k}^{(\pm)}(\boldsymbol r)$ denotes electron magnetic Bloch wavefunction of the $n$th LL at momentum $\boldsymbol k$ in magnetic field along $\pm \hat{z}$ direction.
The expression of the 0LL wavefunction $\Psi_{0,\boldsymbol  k}^{(+)}(\boldsymbol r)$ employed in Eq.~\eqref{chi_mag} is provided in  Appendix \ref{appendix:A}.
 
One solution of functions $\alpha(\boldsymbol r)$ and $\beta(\boldsymbol r)$ in Eq.~\eqref{chi_mag}  is,
\begin{equation}
\begin{aligned}  
\alpha(\boldsymbol r)= & e^{-i  \zeta(\boldsymbol r)/2}\sqrt{[1+n_z(\boldsymbol r)]/2}\big/\lvert\Psi^{(+)}_{0,\boldsymbol\kappa_-}(\boldsymbol r)\rvert, \\
\beta(\boldsymbol r)= & e^{i\zeta(\boldsymbol r)/2}
\sqrt{[1-n_z(\boldsymbol r)]/2}\big/\lvert\Psi^{(+)}_{0,\boldsymbol\kappa_+}(\boldsymbol r)\rvert, \\
\zeta(\boldsymbol r)= & \text{Arg}\{[n_x(\boldsymbol r)+in_y(\boldsymbol r)]\big/[\Psi^{(+)}_{0,\boldsymbol\kappa_-}(\boldsymbol r)]^*\Psi^{(+)}_{0,\boldsymbol\kappa_+}(\boldsymbol r)\}. 
\end{aligned}
\label{abz}
\end{equation}
Here $n_{x,y,z}$ are components of $\boldsymbol{n}$.
Because the zeros of $\lvert\Psi^{(+)}_{0,\boldsymbol\kappa_{\mp}}(\boldsymbol r)\rvert$ and $[\Psi^{(+)}_{0,\boldsymbol\kappa_-}(\boldsymbol r)]^*\Psi^{(+)}_{0,\boldsymbol\kappa_+}(\boldsymbol r)$ are exactly canceled, respectively, by those of $\sqrt{1\pm n_z(\boldsymbol r)}$ and $n_x(\boldsymbol r)+in_y(\boldsymbol r)$, $\alpha(\boldsymbol r)$ and $\beta(\boldsymbol r)$ are periodic functions without singularity. 
Therefore, the spinors $\boldsymbol\chi^{(\pm)}(\bm r)$ obey \textit{magnetic} Bloch's theorem under lattice translation, making $U(\bm r)$ quasiperiodic.
The point group symmetry of $U(\bm r)$ is analyzed in Appendix \ref{appendix:A}.

We now apply $U(\boldsymbol r)$ to the Hamiltonian $H_1$,
\begin{equation}
\begin{aligned}H_2=&U^\dagger(\boldsymbol{r})H_1U(\boldsymbol{r}) \\
=&-\frac{\hbar^2}{2m^*}
\begin{pmatrix}
(\hat{\boldsymbol{k}}+\boldsymbol A_{11})^2
&\hat{\boldsymbol{k}}\boldsymbol A_{12}+\boldsymbol A_{12}\hat{\boldsymbol{k}}\\
\hat{\boldsymbol{k}}\boldsymbol A_{21}+\boldsymbol A_{21}\hat{\boldsymbol{k}}&
(\hat{\boldsymbol{k}}+\boldsymbol A_{22})^2
\end{pmatrix}\\
&+\lvert\boldsymbol \Delta(\boldsymbol{r})
\rvert\sigma_z+[\Delta_0(\boldsymbol{r})-D(\boldsymbol{r})]\sigma_0, 
\end{aligned}
\end{equation}
where $\boldsymbol{A}= -iU^\dagger(\boldsymbol r)\boldsymbol \nabla U(\boldsymbol r)$ is a non-Abelian gauge field generated by the skyrmion field and $D(\mathbf{r})=\frac{\hbar^2}{2m^*}\lvert \boldsymbol A_{12}\rvert^2$ \cite{Nicolas2024Magic,Yu2019Giant}. The field $\boldsymbol A_{ii}$ generates an effective magnetic field $B_i =(\hbar/e)\boldsymbol{\nabla}\times \boldsymbol{A}_{ii}$ with $B_1=+(\hbar/e)b_z$ and $B_2=-(\hbar/e)b_z$. The flux of $B_{1}$  ($B_{2}$) over $\mathcal{A}_0$ is exactly a negative (positive) magnetic flux quantum, according to Eq.~\eqref{Nw}. Therefore, wavefunction $\Phi_{m,\boldsymbol  k}(\boldsymbol r)$ of Hamiltonian $H_2$ for the $m$th band at momentum $\boldsymbol k$ can be expanded using the magnetic Bloch state $\Psi_{n,\boldsymbol  k}^{(\pm)}(\boldsymbol r)$,
\begin{align}
\Phi_{m,\boldsymbol  k}(\boldsymbol r)=  & \sum_{n}[ c_{m,n,\boldsymbol k}^{(+)}\Psi_{n,\boldsymbol  k}^{(-)}(\boldsymbol r),c_{m,n,\boldsymbol k}^{(-)}\Psi_{n,\boldsymbol  k}^{(+)}(\boldsymbol r)]^T\label{C1},
\end{align}
where $c_{m,n,\boldsymbol k}^{(\pm)}$ are expansion coefficients. 

The Bloch state $\varphi_{m,\boldsymbol  k}(\boldsymbol r)$ of the original Hamiltonian $H$ is related to $\Phi_{m,\boldsymbol  k}(\boldsymbol r)$ by unitary transformations,
\begin{equation}
\begin{aligned}
\varphi_{m,\boldsymbol  k}(\boldsymbol r)
=&U_0(\boldsymbol{r})U(\boldsymbol{r})\Phi_{m,\boldsymbol{k}}(\boldsymbol{r})\\
=& \sum_{n}\sum_{s=\pm} c_{m,n,\boldsymbol k}^{(s)}\psi_{n,\boldsymbol  k}^{(s)}(\boldsymbol r).
\end{aligned}    
\label{phidecom}
\end{equation}
Here $\psi_{n,\boldsymbol  k}^{(s)}(\boldsymbol r)=U_0(\boldsymbol{r})\boldsymbol \chi^{(s)}\Psi_{n,\boldsymbol k}^{(-s)}$  are a set of orthonormal and complete Bloch bases, which can be understood as
the LL wavefunction $\Psi_{n,\boldsymbol k}^{(-s)}$ dressed by the spinor $U_0(\boldsymbol{r})\boldsymbol \chi^{(s)}$.
Because $\boldsymbol \chi^{(s)}$ and $\Psi_{n,\boldsymbol k}^{(-s)}$ are magnetic Bloch states under opposite magnetic fields, $\psi_{n,\boldsymbol  k}^{(\pm)}$ satisfy Bloch's theorem. Notably,  $\psi_{n,\boldsymbol  k}^{(\pm)}$ have the same \textit{uniform} quantum geometric tensor as that of $\Psi_{n,\boldsymbol k}^{(\pm)}$, since the spinor $U_{0}(\boldsymbol r)\boldsymbol\chi^{(\pm)}(\boldsymbol r)$ is $\boldsymbol{k}$ independent and normalized at every $\boldsymbol{r}$. Equation~\eqref{phidecom} establishes a LL representation of Bloch states.

To  measure the  overlap between $\varphi_{m,\boldsymbol k}$ and $\psi_{n,\boldsymbol k}^{(s)}$, we define the LL weight,
\begin{align}
W_{m,n}^{(s)}=&\frac{1}{N}\sum_{\boldsymbol k}\lvert c_{m,n,\boldsymbol{k}}^{(s)}\rvert^2,
\end{align}
where $N$ is the number of $\boldsymbol{k}$ points. We focus on the first moir\'e band $\varphi_{1,\boldsymbol k}$, which has a dominant contribution from $\psi_{n,\boldsymbol k}^{(+)}$, since the spinor part of $\psi_{n,\boldsymbol k}^{(+)}$ locally follows the skyrmion field $\boldsymbol{\Delta}(\boldsymbol{r})$. We numerically find that the weight $W_{1,0}^{(+)}$ on the dressed 0LL wavefunction $\psi_{0,\boldsymbol k}^{(+)}$ is sizable ($>0.6$) over a  range of twist angle $\theta \in (2.5^{\circ}, 4.5^{\circ})$, as shown in Fig.~\ref{fig:1}(c). However, $W_{1,0}^{(+)}$ is noticeably below $1$ even at $\theta_m$, because the nonuniform quantum geometry of $\varphi_{1,\boldsymbol k}$ is not captured by $\psi_{0,\boldsymbol k}^{(+)}$.

\section{Variational mapping}
\label{section:4}
To further optimize the weight on the 0LL, we generalize $\psi_{0,\boldsymbol  k}^{(+)}(\boldsymbol r)$ to $\Theta_{\boldsymbol  k}(\boldsymbol r)$,
\begin{align}
\label{wavef1}
\Theta_{\boldsymbol  k}(\boldsymbol r)=\mathcal N_{\boldsymbol  k}U_0(\boldsymbol r)\mathcal{B}(\boldsymbol r)\Psi_{0,\boldsymbol k}^{(-)}(\boldsymbol r),
\end{align}
where $\mathcal{B}(\boldsymbol r)=[\mathcal{B}_1(\boldsymbol r),\mathcal{B}_2(\boldsymbol r)]^T$ has two components and $\mathcal N_{\boldsymbol  k}$ is a normalization factor. Here $\mathcal{B}(\boldsymbol r)$ is not required to be normalized, but $\Theta_{\boldsymbol k}(\boldsymbol r)$ always has a Chern number $\mathcal{C}=1$ and an ideal but nonuniform quantum geometry, $\widetilde g_{\boldsymbol{k}}=\widetilde\Omega_{\boldsymbol{k}}\mathds{1}/2$, where $\widetilde g_{\boldsymbol{k}}$ ($\widetilde\Omega_{\boldsymbol{k}}$) is the quantum metric (Berry curvature) of $\Theta_{\boldsymbol k}(\boldsymbol r)$ \cite{Wang2021Exact}.  We note that $\Theta_{\boldsymbol  k}(\boldsymbol r)$ is also the form of wavefunction for the flatbands in twisted bilayer graphene in the chiral limit \cite{Tarnopolsky2019,Ledwith2020,Wang2021Chiral}.

The advantage of using the generalized 0LL wavefunction $\Theta_{\boldsymbol  k}(\boldsymbol r)$ are threefolds. First, $\mathcal{B}(\boldsymbol r)$ does not need to be an eigenstate of $\bm{\Delta}(\bm r)\cdot \bm{\sigma}$, as the moir\'e wavefunction does not have to follow the skyrmion field exactly.
Second, $\mathcal{B}(\boldsymbol r)$ does not need to be normalized, and the spatial variation of $|\mathcal{B}(\boldsymbol r)|$ can capture the density modulation. 
Third, $\Theta_{\boldsymbol  k}(\boldsymbol r)$ can characterize the nonuniform quantum geometry of the moir\'e band, while $\psi_{0,\boldsymbol  k}^{(+)}$ can not.

We define a new weight
$\widetilde W=\frac{1}{N}\sum_{\boldsymbol k}\lvert \widetilde c_{\boldsymbol{k}}\rvert^2$,
where $\widetilde c_{\boldsymbol{k}}=\langle\varphi_{1,\boldsymbol k}\lvert\Theta_{\boldsymbol k}\rangle$.
We maximize $\widetilde W$ using a variational approach. Taking $\boldsymbol{\chi}^{(+)}(\boldsymbol r)$ as an initial ansatz for  $\mathcal{B}(\boldsymbol r)$, we update $\mathcal{B}(\boldsymbol r)$ step-by-step using the gradient ascend method until convergence, 
\begin{equation}
\label{iter}
\begin{aligned}
\mathrm{Re}[\mathcal{B}_i(\boldsymbol r)]\rightarrow&
\mathrm{Re}[\mathcal{B}_i(\boldsymbol r)]+\xi \frac{\delta\widetilde W}{\delta \mathrm{Re}[\mathcal{B}_i(\boldsymbol r)]},\\
\mathrm{Im}[\mathcal{B}_i(\boldsymbol r)]\rightarrow&
\mathrm{Im}[\mathcal{B}_i(\boldsymbol r)]+\xi \frac{\delta\widetilde W}{\delta \mathrm{Im}[\mathcal{B}_i(\boldsymbol r)]},\\
\end{aligned}    
\end{equation}
where $\xi$ is a positive parameter (see Appendix \ref{appendix:B} for details).

The maximized $\widetilde W$ as a function of $\theta$ is shown in Fig.~\ref{fig:1}(c), which has a significant increase compared to $W_{1,0}^{(+)}$ particularly around $\theta_m$. For $\theta \in (2.5^\circ,3.7^\circ)$, $\widetilde W$ exceeds 0.9 and achieves a maximum of 0.95 at $\theta_m$.  The momentum dependence of $\lvert \widetilde c_{\boldsymbol{k}}\rvert$ at $\theta_m$ is shown in Fig.~\ref{fig:1}(d), which is larger than $0.94$ for every $\boldsymbol{k}$ and reaches nearly $1$ at the $\boldsymbol\gamma$ point, indicating a high overlap between $\varphi_{1,\boldsymbol k}$ and $\Theta_{\boldsymbol k}$. The spatial variation of $\lvert \mathcal{B}(\boldsymbol r) \rvert$ is shown in Fig.~\ref{fig:1}(e), where the maximal positions form an effective honeycomb lattice, consistent with the density distribution in the first band \cite{Qiu2023}.
Moreover, the quantum geometry fluctuation in $\varphi_{1,\boldsymbol{k}}(\boldsymbol r)$ is qualitatively captured by $\Theta_{\boldsymbol  k}(\boldsymbol r)$, as illustrated in Fig.~\ref{fig:2}. We plot the vector field $\widetilde{\boldsymbol{n}}(\boldsymbol{r})$ defined as $[\mathcal{B}(\boldsymbol r)]^\dagger \bm{\sigma} \mathcal{B}(\boldsymbol r)$ in Fig.~\ref{fig:3}(b) at $\theta_m$. Compared to $\boldsymbol{n}(\boldsymbol{r})=[\chi^{(+)}(\boldsymbol r)]^\dagger \bm{\sigma} \chi^{(+)}(\boldsymbol r)$, $\widetilde{\boldsymbol{n}}(\boldsymbol{r})$ also forms a skyrmion lattice but has spatially varying magnitude.

\begin{figure}[t]
    \includegraphics[width=1.\columnwidth]{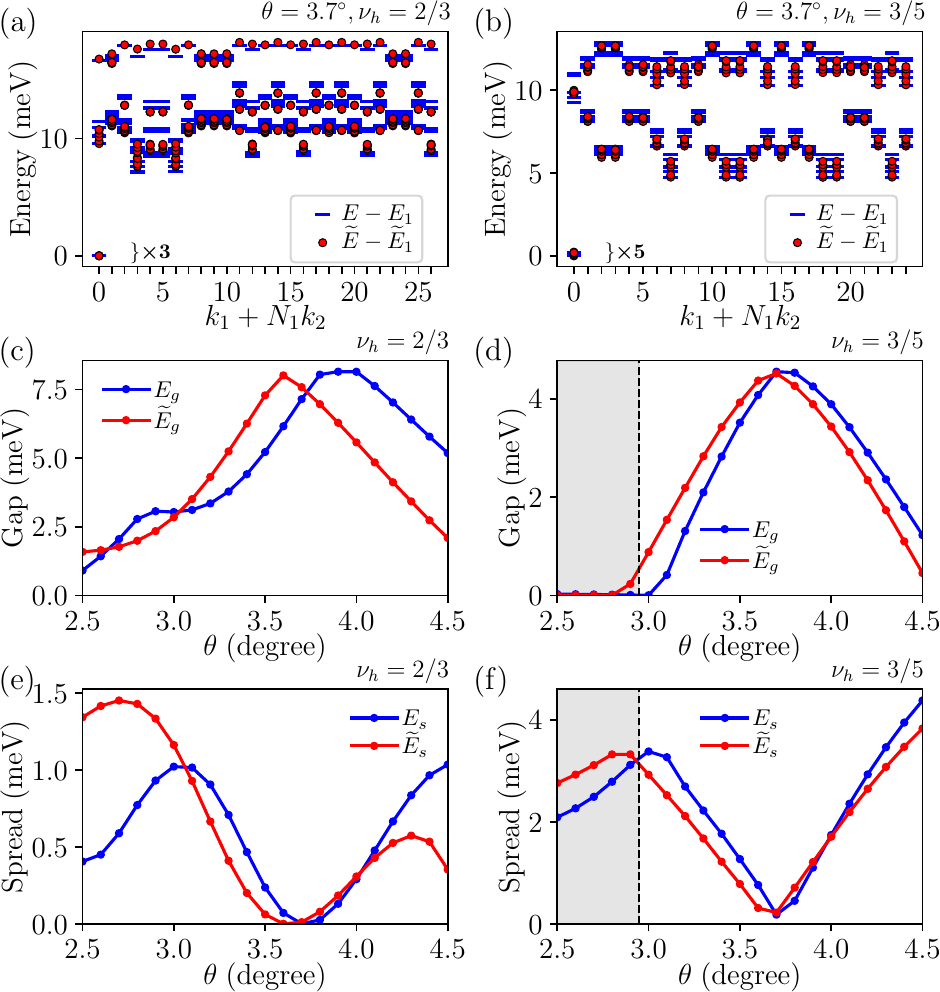}
    \caption{(a), (b) ED spectra of the original model (blue lines) and variational model (red dots) at $\nu_h=2/3$ and $3/5$ for $\theta=3.7^\circ$.  (c)-(f) Energy gap $E_g$ and spread $E_s$  as functions of $\theta$ at $\nu_h=2/3$ and $3/5$. Blue (red) lines present results from the original (variational) model. The FCIs are absent in the gray regions of (d) and (f), where $E_g$ is zero.}
    \label{fig:4}
\end{figure}

\section{Fractional Chern Insulators} 
\label{section:5}
A remarkable property of the generalized 0LL with $\Theta_{\boldsymbol k}(\boldsymbol r)$ is that it allows construction of trial wavefunction for FCIs \cite{Wang2021Exact,Ledwith2020},
$\Phi_F=\Psi_F\prod_{i} U_0 (\boldsymbol{r}_i) \mathcal{B}(\boldsymbol{r}_i)$,
where $\boldsymbol{r}_i$ is the position of the $i$th electron and $\Psi_F$ represents the fractional quantum Hall states (or composite Fermi liquid states) in the 0LL.
Here $\Phi_F$ can be the exact ground state of certain short-range repulsive interactions \cite{Wang2021Exact}.
The high overlap between $\varphi_{1,\boldsymbol{k}}$ and $\Theta_{\boldsymbol k}(\boldsymbol r)$ provides a rational explanation for FCIs observed in $t$MoTe$_2$.

\begin{figure*}
    \includegraphics[width=2.\columnwidth]{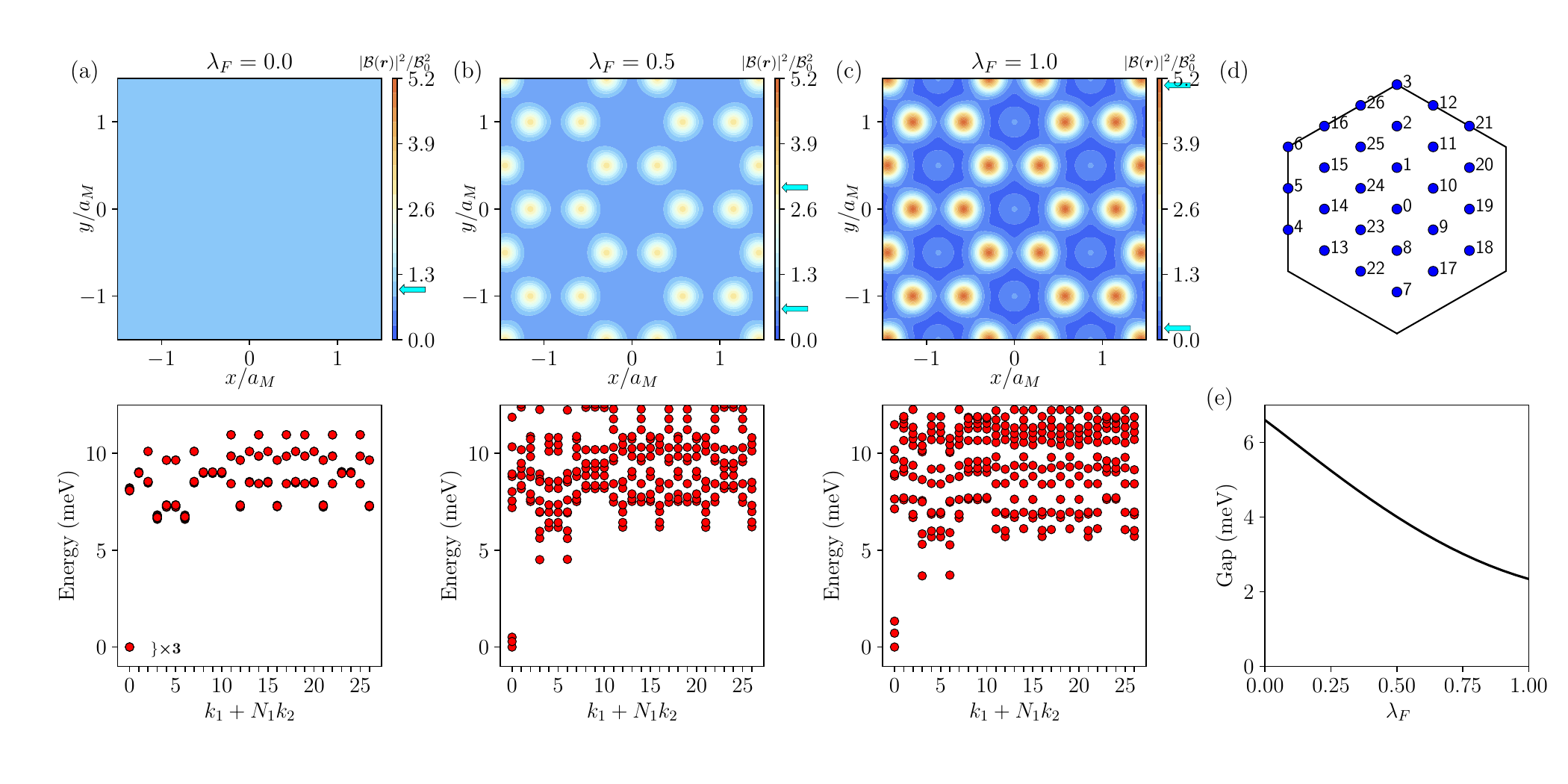}     \caption{ (a-c) $\lvert \mathcal{B}(\boldsymbol{r},\lambda_F)\rvert^2$ scaled by its spatial average (upper panel) and energy spectra (lower panel) for $\lambda_F=0,0.5,1$ at $\nu_h=2/3$ and $\theta=2.9^\circ$. (d) The 27-unit-cell momentum clusters used in the ED calculations. The integer numbers are the momentum label $k_1+N_1 k_2$. (e) Energy gap as a function of $\lambda_F$ at $\nu_h=2/3$ and $\theta=2.9^\circ$.} 
    \label{fig:5}
\end{figure*}

We quantitatively investigate FCIs by studying a many-body Hamiltonian with Coulomb interactions, 
\begin{equation}
\begin{aligned}
\hat{\mathcal{H}} = \sum_{\boldsymbol{k}}(-\mathcal{E}_{\boldsymbol{k}})b^\dagger_{\boldsymbol{k}}b_{\boldsymbol{k}}+\sum_{\boldsymbol{k_1k_2k_3k_4}}V_{\boldsymbol{k_1k_2k_3k_4}}
b^\dagger_{\boldsymbol{k_1}}b^\dagger_{\boldsymbol{k_2}}
b_{\boldsymbol{k_3}}b_{\boldsymbol{k_4}},
\end{aligned}   
\end{equation}
where we only keep states in the first moir\'e band at $+K$ valley assuming spontaneous valley polarization, $b_{\bm k }^{\dagger}$ ($b_{\bm k }$) is the  creation (annihilation) operator in the \textit{hole} basis, and $-\mathcal{E}_{\boldsymbol{k}}$ is the hole single-particle energy. The interaction matrix element $V_{\boldsymbol{k_1k_2k_3k_4}}$ is 
\begin{equation}
\begin{aligned}
    V_{\boldsymbol{k_1k_2k_3k_4}}=&\frac{1}{2\mathcal{A}}\sum_{\boldsymbol{q}}V_{\boldsymbol{q}}
    M_{\boldsymbol{k_1}\boldsymbol{k_4}}(\boldsymbol{q})
    M_{\boldsymbol{k_2}\boldsymbol{k_3}}(\boldsymbol{-q}),\\
 M_{\boldsymbol{k}\boldsymbol{k'}}(\boldsymbol{q})
=&\int d\boldsymbol{r}\,e^{i\boldsymbol{q}\cdot \boldsymbol{r}}[f_{\boldsymbol{k}}(\boldsymbol{r})]^*f_{\boldsymbol{k'}}(\boldsymbol{r}),
\end{aligned}
\end{equation}
where $\mathcal{A}$ is the system area, $V_{\boldsymbol{q}}=2\pi e^2\tanh{\lvert\boldsymbol{q}\rvert d}/(\epsilon\lvert\boldsymbol{q}\rvert)$ is gate screened Coulomb interaction, $d$ is the gate-to-sample distance, and $\epsilon$ is the dielectric constant. We set $d=100
\;\rm{nm}$ and $\epsilon=5$.  In $M_{\boldsymbol{k}\boldsymbol{k'}}(\boldsymbol{q})$, 
$f_{\boldsymbol{k}}(\boldsymbol{r})$ is $[\varphi_{1,\boldsymbol k}(\boldsymbol r)]^*$ or $[\Theta_{\boldsymbol k}(\boldsymbol r)]^*$, corresponding to the original model and variational model, respectively.

We compare ED spectra of the above two models at $\theta=3.7^{\circ}$ in Fig.~\ref{fig:4}(a) [\ref{fig:4}(b)] for $\nu_h= 2/3$ ($3/5$), which are obtained using clusters with 27 (25) unit cells. The three (five) quasi-degenerate ground states in the zero momentum sector indicate the presence of the FCIs for both models at $\nu_h= 2/3$ ($3/5$) \cite{Regnault2011Fractional,Reddy2023Fractional,yu2024fractional}. Moreover, the excited-state spectra of the original model are approximately reproduced by those of the variational model, as shown in Figs.~\ref{fig:4}(a) and \ref{fig:4}(b).

Figure~\ref{fig:4} also plots the charge-neutral  gap $E_g=E_4-E_3$ $(E_6-E_5)$ and ground-state energy spread $E_s=E_3-E_1$ $(E_5-E_1)$ as functions of $\theta$ for $\nu_h= 2/3$ ($3/5$), where $E_n$ is the $n$th lowest energy in the ED spectra.  At $\nu_h= 2/3$, $E_g$ is maximum at $\theta=3.9^{\circ}$ ($3.6^{\circ}$) in the original (variational) model and is finite for  $\theta \in (2.5^\circ, 4.5^\circ)$, indicating a robust FCI phase; $E_s$ is vanishingly small around $\theta=3.7^{\circ}$ in both models. The two models also compare quantitatively well at $\nu_h= 3/5$ and share the following features, (1) $E_g$ is maximum and $E_s$ is minimum at $\theta=3.7^{\circ}$; (2) $E_g$ is finite only for $\theta \gtrsim 3^{\circ}$, indicating a narrower range of FCI phase than $\nu_h=2/3$.

An intriguing observation is that the strongest FCI measured by the maximum $E_g$ or minimum $E_s$ does not occur at $\theta_m$. This can be understood based on the variational wavefunction $\Theta_{\boldsymbol k}(\boldsymbol r)$. The spatial variation of $\lvert \mathcal{B}(\boldsymbol r) \rvert$ tends to drive phase transition from FCI to other competing states, weakening the FCI phase if not destroyed. As $\theta$ increases,  the effect of moir\'e potential confinement decreases, reducing the spatial variation of $\lvert \mathcal{B}(\boldsymbol r) \rvert$. Therefore, the gap $E_g$  at $\nu_h= 2/3$ increases as $\theta$ increases away from $\theta_m$, where the effect of bandwidth is minimal. 
To place this physical argument at a more quantitative level, we expand the periodic function $\left|\mathcal{B}\left(\boldsymbol{r}\right)\right|^2$ in terms of Fourier series
\begin{equation}
\begin{aligned}
\lvert \mathcal B(\boldsymbol {r})\rvert^2=&\mathcal{B}_0^2 (1+\sum_{\boldsymbol g\ne\boldsymbol {0}}F_{\boldsymbol g}e^{i\boldsymbol{g}\cdot\boldsymbol{r}}),\\
\end{aligned}    
\end{equation}
where $\mathcal{B}_0^2$ is the spatial average of $\left|\mathcal{B}\left(\boldsymbol{r}\right)\right|^2$ and $\boldsymbol{g}$ represents the moiré reciprocal lattice vectors. $\mathcal{B}_0^2$ can be absorbed into the normalization factor, so that the Fourier coefficients $F_{\boldsymbol{g}}$ quantifies the spatial variation and fully determines the interaction matrix element. We note that $\mathcal{B}\left(\boldsymbol{r}\right)$ is quasiperiodic with magnetic translational symmetry, but $|\mathcal{B}\left(\boldsymbol{r}\right)|$ is periodic.

To clearly illustrate the effect of the\ spatial variation in $\left|\mathcal{B}\left(\boldsymbol{r}\right)\right|$, we introduce a parameter $\lambda_F$ to control the variation as follows,
\begin{equation}
\begin{aligned}
\lvert \mathcal B(\boldsymbol r,\lambda_F)\rvert^2=&\mathcal{B}_0^2 (1+\lambda_F\sum_{\boldsymbol g\ne\boldsymbol{0}}F_{\boldsymbol g}e^{i\boldsymbol{g}\cdot\boldsymbol{r}}).   
\end{aligned}    
\end{equation}
In the limit of $\lambda_F=0$, $\lvert \mathcal B(\boldsymbol r,\lambda_F)\rvert^2$ is spatially uniform, and the interaction matrix elements become the same as those of lowest Landau level; therefore, the physics of lowest Landau level is fully recovered if the single-particle bandwidth can be further neglected. In the opposite limit of $\lambda_F=1$, $\lvert \mathcal B(\boldsymbol r,\lambda_F)\rvert^2$ reduces to $\lvert \mathcal B(\boldsymbol r)\rvert^2$ of the problem under study. The parameter $\lambda_F$ provides a convenient tuning knob in examining the evolution of the spectrum as the spatial variation of $\left|\mathcal{B}\left(\boldsymbol{r}, \lambda_F\right)\right|^2$ changes. 

\begin{figure}[t]
    \includegraphics[width=1.\columnwidth]{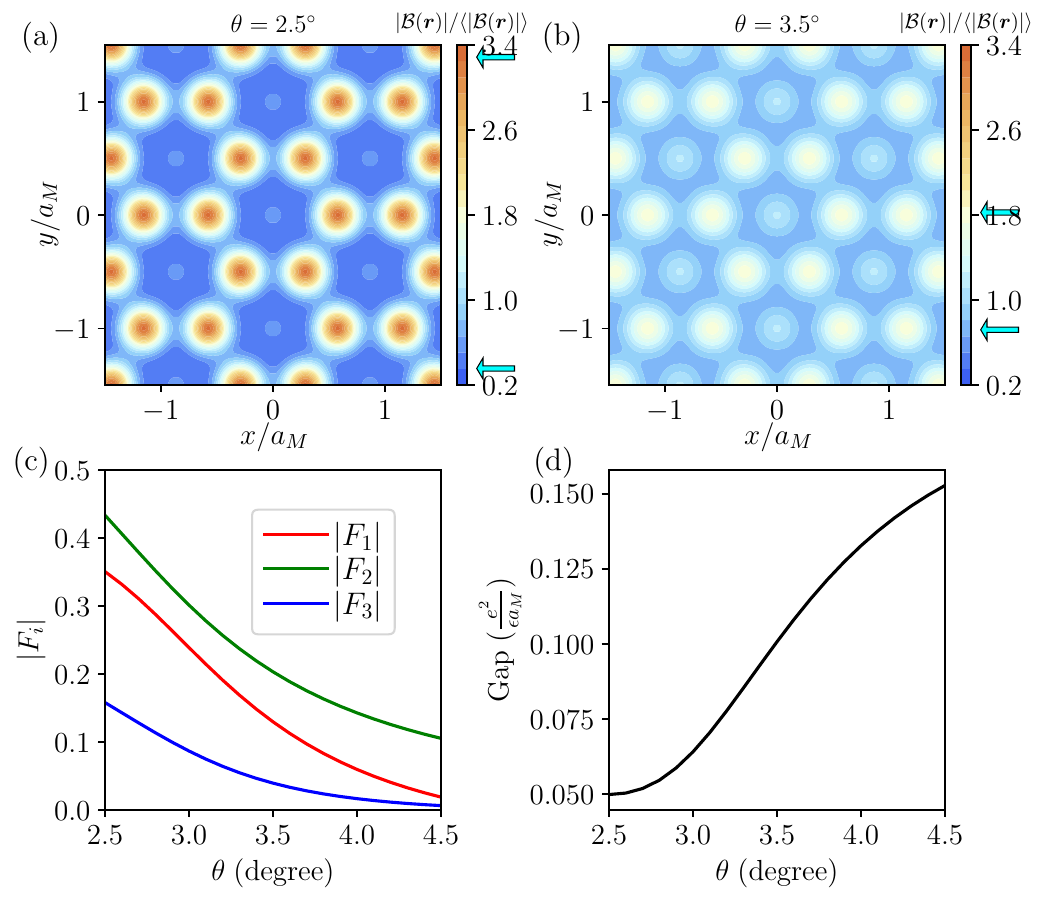}
     \caption{(a-b) Map of $\lvert \mathcal{B}(\boldsymbol{r})\rvert$ scaled by its spatial average at $\theta=2.5^{\circ}$ and $3.5^{\circ}$. (c) Absolute value of Fourier components $\lvert F_i\rvert$ of $\lvert \mathcal B(\boldsymbol{r})\rvert^2$. (d) Gap as a function of $\theta$ at $\nu_h=2/3$. Here we use $e^2/(\epsilon a_M)$ as the unit of energy, and the single-particle bandwidth is set to 0.} 
    \label{fig:6}
\end{figure}

In Figs.~\ref{fig:5}(a)-(c), we show $\lvert \mathcal B(\boldsymbol r,\lambda_F)\rvert^2$ scaled by its spatial average for $\lambda_F$=0, 0.5, and 1 at $\theta=2.9^\circ$ in the upper panels, and the corresponding ED energy spectra for the filling factor $\nu_h=2/3$ in the lower panels. As $\lambda_F$ increases, the fluctuation of $\lvert \mathcal B(\boldsymbol r,\lambda_F)\rvert^2$ increases, and the single-particle wavefunction has larger weights on the maximal points of $\lvert \mathcal B(\boldsymbol r,\lambda_F)\rvert^2$. 
For $\lambda_F=0$, the energy spectra are essentially the same as those for lowest Landau level at $\nu_h=2/3$, and host threefold degenerate ground states at the zero momentum sector [momentum cluster shown in Fig.~\ref{fig:5}(d)] that are separated by an energy gap from excited states, a hallmark of the fractional state. As $\lambda_F$ increases, the degeneracy in the ground state manifold is gradually broken. Meanwhile, excited states at momentum sectors 3 and 6, which correspond to the Brillouin zone corners as depicted in Fig.~\ref{fig:5}(d), decrease in energy relative to the ground state. Therefore, the gap between the ground state manifold and excited states decreases as the spatial variation of $\lvert \mathcal B(\boldsymbol r,\lambda_F)\rvert^2$ increases, as shown in Fig.~\ref{fig:5}(e).

The reason behind this numerical observation can be understood as follows. With the increase of the spatial variation of $\lvert \mathcal B(\boldsymbol r,\lambda_F)\rvert$, the wavefunction are more concentrated near the maxima positions of $\lvert \mathcal B(\boldsymbol r,\lambda_F)\rvert$. The more localized nature of the wavefunction has a tendency to drive phase transition from FCI state to charge density wave (CDW) state. For the results presented in Fig.~\ref{fig:5}, the ground state remains in the FCI phase, but excited states at momenta of the Brillouin zone corners come down in energy, which are precursors of CDW phase. This explains the decrease of the gap with the increase of $\lambda_F$.

We now consider the twist angle dependence and fix $\lambda_F=1$.
We show the $\theta$ dependence of $\lvert \mathcal B(\boldsymbol r)\rvert$ in Fig.~\ref{fig:6}(a-b) and the evolution of $\lvert F_i\rvert$ as a function of  $\theta$ in Fig.~\ref{fig:6}(c), where $F_i$ denotes the value of the Fourier coefficients of $\lvert \mathcal B(\boldsymbol{r})\rvert^2$ in the $i$th momentum shell. As $\theta$ increases, $\lvert F_i\rvert$ decreases, which results from the weaker moiré confinement at larger twist angles. In Fig.~\ref{fig:6}(d), we show the charge neutral gap as a function of $\theta$ with the bandwidth artificially set to 0, where the gap increases monotonically with increasing $\theta$. When the actual bandwidth is turned on, we find that the gap still increases as $\theta$ increases away from the magic angle, but reaches a maximal value at $\theta$ around $3.6^\circ$, and then decreases. The decrease in the gap for $\theta>3.6^\circ$ can be attributed to the increase of the bandwidth. 
Overall, the FCI phase is not the strongest at the magic angle because the spatial variation of $\lvert \mathcal B(\boldsymbol{r})\rvert$ is also strong here. The variational wavefunction provides a convenient way to analyze this effect.

\section{Further application}
\label{section:6}

In $t$MoTe$_2$, the layer pseudospin skyrmion field can be identified in the Hamiltonian, which leads to an emergent magnetic field and allows us to map Chern bands to Landau levels. Here we generalize our method to the generic case without the emergent magnetic field. We consider a model proposed in Ref.~\cite{tan2024designing}. The Hamiltonian is written as
\begin{equation}
\begin{aligned}
H_{\tau}(\mathbf{k})=
\begin{pmatrix}
\alpha_{1}|\mathbf{k}|^{2}+\frac{\delta}{2}+V(\mathbf{r}) & v\left(\tau k_{x}-i k_{y}\right) \\
v\left(\tau k_{x}+i k_{y}\right) & -\alpha_{2}|\mathbf{k}|^{2}-\frac{\delta}{2}+V(\mathbf{r}),
\end{pmatrix}
\end{aligned}    
\end{equation}
which describes a narrow gap semiconductor in a superlattice potential  $V(\boldsymbol{r})=2V_0\sum_{n=1}^3\cos{(\boldsymbol{g}_n\cdot\boldsymbol{r}+\phi)}$ with $\boldsymbol{g}_n=g(-\cos{\frac{2\pi n}{3}},\sin{\frac{2\pi n}{3} })$, $g=4\pi/\sqrt{3}a$, and $a$ being the superlattice period. Here $\tau=\pm$ is the spin index. This model can host a Chern band that nearly saturates the trace inequality $T\ge0$ in certain parameter regimes, but no skyrmion field or emergent magnetic field can be identified in the Hamiltonian. We take $\phi=\pi,\alpha_1=\alpha_2=\alpha$, $g/k_D =2$ and $V_0/E_D=1.5$ in our calculation, where $k_D=\nu/\alpha$ and $E_D=\nu^2/\alpha$ are characteristic momentum and energy, respectively. 

\begin{figure}[t]
\includegraphics[width=1.\columnwidth]{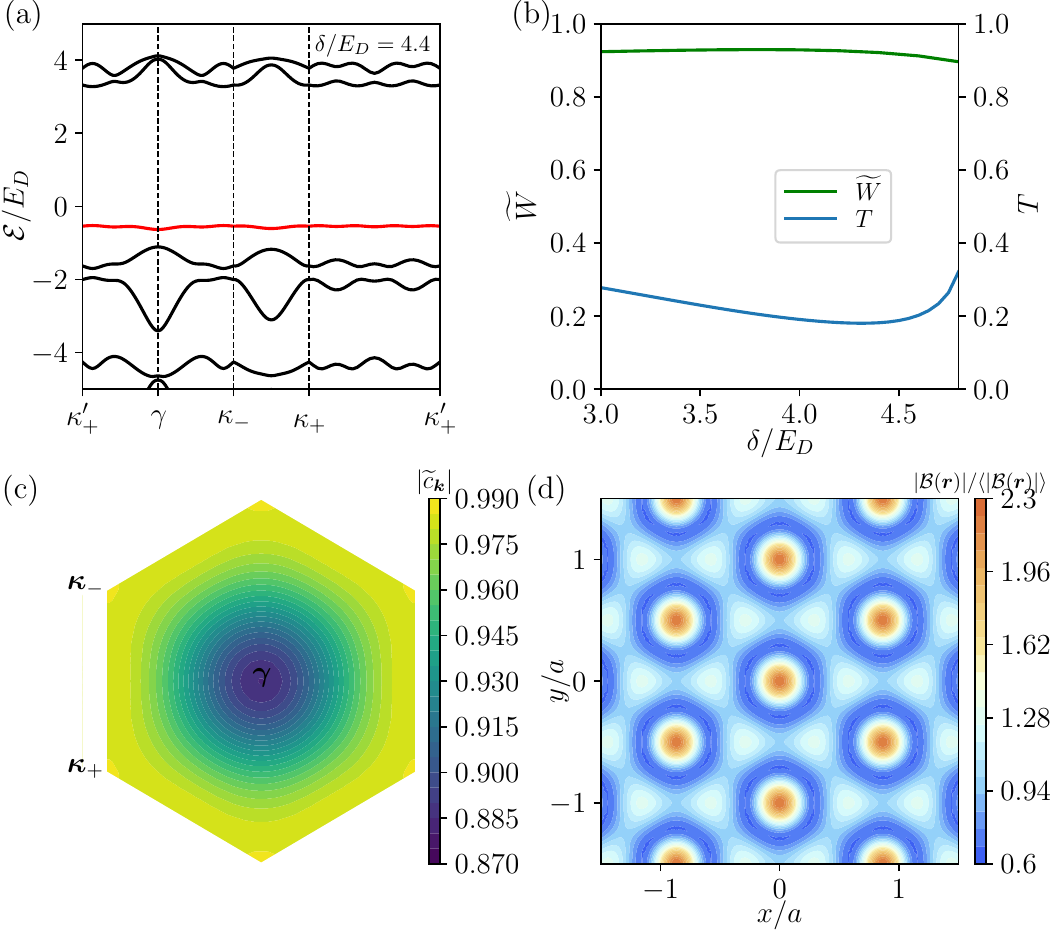}
     \caption{(a) Band structure of the model Hamiltonian $H_\tau(\boldsymbol{k})$ at $\delta/E_D=4.4$ and $\tau=-$. The first conduction band is plotted in red line. (b) Landau-level weight $\widetilde{W}$ and deviation from ideal quantum geometry $T$ of the first conduction band as a function of $\delta/E_D$. (c) Overlap  $\lvert \widetilde{c_{\boldsymbol{k}}}\rvert$ in the Brillouin zone. (d) Map of $\lvert \mathcal B(\boldsymbol{r})\rvert$ scaled by its spatial average. $\delta/E_D=4.4$ in (c) and (d).}
    \label{fig:7}
\end{figure}

Figure~\ref{fig:7}(a) shows the band structure at $\tau=-$ and $\delta/E_D=4.4$, where the first conduction band is energetically isolated from other bands and nearly flat with a Chern number of +1. We further show the deviation $T$ from the ideal quantum geometry in Fig.~\ref{fig:7}(b) for $\delta/E_D\in[3.0,4.8]$, where $T$ can be as small as 0.18. The small value of $T$ indicates the resemblance of the Chern band with the generalized 0LL, of which the wavefunction $\Theta_{\boldsymbol{k}}(\boldsymbol{r})$ takes the same form as in Eq.~\eqref{wavef1}.
For this case, $U_0(\bm r)$ is taken to be an identity matrix and  $\mathcal B(\boldsymbol{r})$ also has two components. We again quantify the resemblance by defining the Landau Level weight $\widetilde W=\frac{1}{N}\sum_{\boldsymbol{k}}\lvert \widetilde c_{\boldsymbol{k}}\rvert^2$, where $\widetilde c_{\boldsymbol{k}}=\langle{\varphi_{\boldsymbol{k}}}\mid \Theta_{\boldsymbol{k}}\rangle$ and $\varphi_{\boldsymbol{k}}$ is the two-component wavefunction for the first conduction band of $H_\tau(\boldsymbol{k})$ with $\tau=-$. We take the following  initial ansatz for $\mathcal B(\boldsymbol{r})$,
\begin{equation}
\begin{aligned}
\mathcal B(\boldsymbol{r})\rightarrow
&\frac{\varphi_{\boldsymbol\gamma}(\boldsymbol r)}{|\varphi_{\boldsymbol\gamma}(\boldsymbol r)|^2}\Psi_{0,\boldsymbol\gamma}^{(+)}(\boldsymbol r),
\end{aligned}    
\end{equation}
where $\varphi_{\boldsymbol\gamma}(\boldsymbol r)$ is the wavefunction of the first conduction band at the $\boldsymbol\gamma$ point.
With this ansatz, we update $\mathcal B(\boldsymbol{r})$ step-by-step by the gradient ascend method until convergence. As presented in Fig.~\ref{fig:7}(b), the optimized $\widetilde W$ is over 0.9 for $\delta/E_D\in[3.0,4.8]$. In Fig.~\ref{fig:7}(c), we plot $\lvert\widetilde c_{\boldsymbol{k}}\rvert$ over the Brillouin zone at $\delta/E_D=4.4$, where $\lvert\widetilde c_{\boldsymbol{k}}\rvert$ is nearly 1 at the Brillouin zone corners. We further plot $\lvert\mathcal B(\boldsymbol{r})\rvert$ in Fig.~\ref{fig:7}(d), which is a continuous function of $\bm{r}$ and respects the symmetry of the system. This example demonstrates that our method is general in mapping the Chern band to generalized Landau levels.

\section{Discussion}
\label{section:7}
Small values of the deviation $T$ from the ideal quantum geometry have often been interpreted as an indication of the resemblance between Chern bands and the generalized 0LL, but the exact form of the latter is unknown (i.e., unknown $\mathcal{B}(\boldsymbol r)$). Our work provides a variation approach to determine the optimized $\mathcal{B}(\boldsymbol r)$ and quantify the resemblance. 
In $t$MoTe$_2$, we show that the variational wavefunction $\Theta_{\boldsymbol k}(\boldsymbol r)$ successfully captures both the single-particle and many-body physics about the formation and properties of FCIs.
We expect applications of our theory in several directions. 
First, the wavefunction $\Phi_F$  can be generalized to excited states of fractionalized quasihole and quasielectron \cite{Laughlin1983}. After $\Theta_{\boldsymbol k}(\boldsymbol r)$ is determined using the variational approach, properties of anyon excitations in FCIs can be explored based on many-body wavefunctions.
Second, the variational mapping between moir\'e bands and higher-index LL can be performed using our approach combined with the recently proposed theory of generalized LL \cite{liu2024theory}. Several theoretical works reported non-Abelian fractional states at half-filling of the second moir\'e band \cite{reddy2024nonabelian,xu2024multiple,ahn2024landau,wang2024higher}, whose connection to the first LL can be worked out.
Finally, FCIs at zero magnetic field have also been observed in a rhombohedral pentalayer graphene–hBN moir\'e superlattice \cite{lu2024fractional} and theoretically proposed in various models \cite{Ghorashi2023Topological,Wan2023Topological,Qiang2023Untwisting,tan2024designing}. Our mapping strategy can be readily applied to different systems to reveal the properties of FCIs.

\section{Acknowledgments}
We thank Jie Wang and Zhao Liu for their valuable discussions. This work is supported by National Natural Science Foundation of China (Grant No. 12274333), and National Key Research and Development Program of China (Grants No. 2021YFA1401300 and  No. 2022YFA1402401). The numerical calculations in this paper have been performed on the supercomputing system in the Supercomputing Center of Wuhan University.

\appendix
\section{Symmetry}
\label{appendix:A}
This appendix presents the symmetry properties of different quantities. We study both point group symmetries and translational symmetry. The translation symmetry is generated by primitive lattice vectors $\boldsymbol{a}_{1,2}=(\pm\frac{\sqrt{
3}}{2},\frac{1}{2})a_M$  of the moir\'e triangular lattice. 

We first study the symmetry properties of magnetic Bloch wavefunctions at the zeroth Landau level, which takes the following form under symmetric gauge,
 \begin{equation}
 \begin{aligned}
     \Psi_{0,\boldsymbol k}^{(-)}(\boldsymbol r)=&\frac{1}{S_{\boldsymbol{k} }\ell}\sigma(z+iz_{\boldsymbol k}\ell^2 )e^{-\frac{1}{4}\lvert z_{\boldsymbol k}\rvert^2\ell^2-\frac{1}{4}\lvert z\rvert^2\ell^{-2}+\frac{i}{2}z_{\boldsymbol k}^*z},\\
     \Psi_{0,\boldsymbol k}^{(+)}(\boldsymbol r)=&[\Psi_{0,-\boldsymbol k}^{(-)}(\boldsymbol r)]^*.
 \end{aligned}     
 \end{equation}
where $z = x + iy, z_{\boldsymbol{k}} = k_x + ik_y$ , $S_{\boldsymbol{k}}$ is a normalization factor, $\ell = \sqrt{\mathcal{A}_0/(2\pi)}$, and $\mathcal{A}_0$ is the area of the (magnetic) unit cell. The modified Weierstrass sigma function $\sigma(z)$ is \cite{Haldane2018modular}
\begin{align}
\label{sigma}
\sigma(z)= ze^{\frac{\eta_1z^2}{z_1}}\frac{\mathcal{\theta}_1(u\mid\tau)}{u\mathcal{\theta}_1'(0\mid\tau)},
\end{align}
where $\mathcal{\theta}_1(u\mid\tau)$ is the Jacobi theta function, $u=\pi z/z_1$, $\eta_1=z_1^*/(4\ell^2)$, $\tau=z_2/z_1$, and $z_j=a_{j,x}+ia_{j,y}$. Expression of $\mathcal{\theta}_1(u\mid\tau)$ is
\begin{align}
\mathcal{\theta}_{1}(u\mid\tau)=-\sum_{n=-\infty}^{+\infty}e^{i\pi\tau(n+\frac{1}{2})^2}e^{2\pi i(n+1/2)(u+1/2)}.
\end{align}

The magnetic Bloch wavefunction satisfies the following magnetic translational symmetry in real space,
\begin{align}
\label{s0}
\Psi_{0,\boldsymbol k}^{(s)}(\boldsymbol r+\boldsymbol{a}_i)
=-e^{-i\frac{1}{2\ell^2}s\boldsymbol {a}_i\times \boldsymbol r}e^{i\boldsymbol k\cdot \boldsymbol {a}_i}\Psi^{(s)}_{0,\boldsymbol k}(\boldsymbol r).
\end{align}
In momentum space, the magnetic Bloch wavefunction satisfies a quasi-periodic boundary condition,
\begin{align}
\label{boundmom}
\Psi_{0,\boldsymbol k+\boldsymbol{b}_i}^{(s)}(\boldsymbol r)
=-e^{-i\frac{\ell^2}{2}s\boldsymbol{b}_i\times \boldsymbol k}\Psi^{(s)}_{0,\boldsymbol k}(\boldsymbol r),
\end{align}
where $\boldsymbol{b}_{1,2}=\frac{4\pi}{\sqrt{3}a_M}(\pm\frac{1}{2},\frac{\sqrt{3}}{2})$ are primitive reciprocal lattice vectors and $\boldsymbol{a}_{i}\cdot \boldsymbol{b}_{j}=2\pi \delta_{ij}$.

It is also useful to consider the original definition of Weierstrass sigma function,
\begin{equation}
\label{sigma2}
\begin{aligned}
\sigma(z)= & z\prod_{w\in\Lambda}(1-\frac{z}{w})\exp(\frac{z}{w}+\frac{z^2}{2w^2}),
\end{aligned}
\end{equation}
where $\Lambda= \{mz_1+nz_2\mid m\in \mathcal{Z},n\in \mathcal{Z},(m,n)\ne(0,0)\}$. It is known that two forms in Eqs.~\eqref{sigma} and \eqref{sigma2} are equivalent in a triangular lattice system \cite{Haldane2018modular}. Based on Eq.~\eqref{sigma2}, we derive the point group symmetry of magnetic Bloch wavefunction at three high symmetry momenta $\boldsymbol{\gamma}=(0,0),\boldsymbol{\kappa}_+=\frac{4\pi}{3 a_M}(-\frac{\sqrt{3}}{2},-\frac{1}{2})$,  and $\boldsymbol{\kappa}_-=\frac{4\pi}{3 a_M}(-\frac{\sqrt{3}}{2},+\frac{1}{2})$.
\begin{equation}
\label{s1}
\begin{aligned} 
 &\Psi^{(s)}_{0,\boldsymbol \gamma}(\hat R_{3z}\boldsymbol r)=e^{-i\frac{2}{3}\pi s}\Psi^{(s)}_{0,\boldsymbol \gamma}(\boldsymbol r),\\
  &\Psi^{(s)}_{0,\boldsymbol \gamma}(\hat{R}_{2z}\boldsymbol r)=-\Psi^{(s)}_{0,\boldsymbol \gamma}(\boldsymbol r),\\
 &\Psi^{(s)}_{0,\boldsymbol \gamma}(\hat M_{x}\boldsymbol r)=-[\Psi^{(s)}_{0,\boldsymbol \gamma}(\boldsymbol r)]^*,\\
 &\Psi^{(s)}_{0,\boldsymbol{\kappa}_\pm}(\hat R_{3z}\boldsymbol r)=\Psi^{(s)}_{0,\boldsymbol{\kappa}_\pm}(\boldsymbol r),\\
  &\Psi^{(s)}_{0,\boldsymbol {\kappa}_\pm}(\hat{R}_{2z}\boldsymbol r)=-e^{\pm i\frac{2}{3}\pi s}\Psi^{(s)}_{0,\boldsymbol {\kappa}_\mp}(\boldsymbol r),\\ 
  & \Psi^{(s)}_{0,\boldsymbol {\kappa}_\pm}(\hat M_{x}\boldsymbol r)=-[\Psi^{(s)}_{0,\boldsymbol {\kappa}_\mp}(\boldsymbol r)]^*,\\
 \end{aligned}
\end{equation}
where $\hat{R}_{nz}$ represents the $n$-fold rotation around the $z$-axis and $\hat M_x$ is the mirror operation that flips  $x$ to $-x$.

\begin{figure}[t]
    \includegraphics[width=1.\columnwidth]{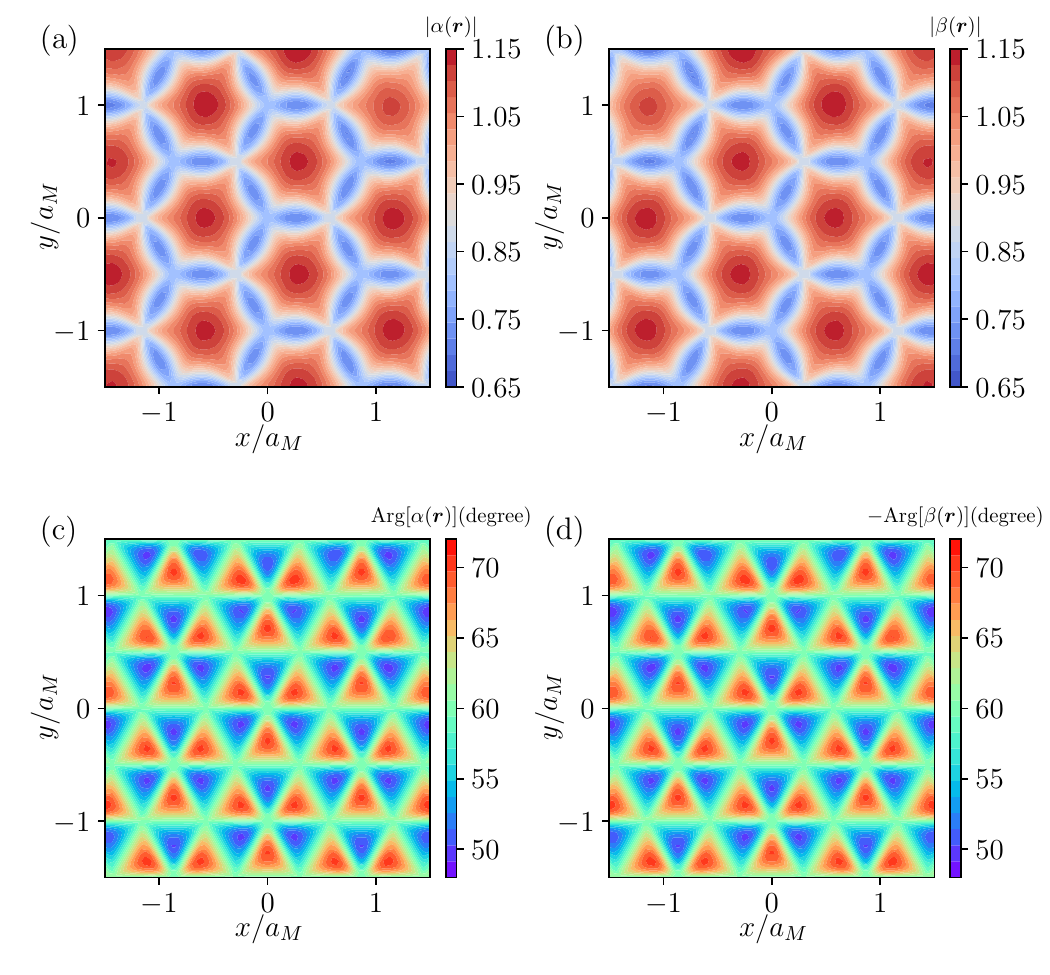}
    \caption{(a), (b) Spatial distribution of  amplitudes $\lvert\alpha(\boldsymbol{r})\rvert$ and $\lvert\beta(\boldsymbol{r})\rvert$. (c), (d) Spatial distribution of phases $\mathrm{Arg}[\alpha(\boldsymbol{r})]$ and $-\mathrm{Arg}[\beta(\boldsymbol{r})]$.} 
    \label{fig:8}
\end{figure}

We turn to the moir\'e Hamiltonian $H$ of $t$MoTe$_2$, which is invariant under $\hat{C}_{3z}$, $\hat{C}_{2y}\hat{\mathcal{T}}$, and an effective inversion symmetry $\hat{\mathcal{I}}$, where $\hat{C}_{nj}$ represents $n$-fold rotational symmetry around $j$ axis and  $\hat{\mathcal{T}}$ is the time-reversal symmetry. Here $\hat{C}_{2y}$ is a combination of $\hat{M}_x$ and layer exchange $\sigma_x$, where $\sigma_x$ is the $x$ Pauli matrix in the layer pseudospin space. Both $\hat{C}_{3z}$ and $\hat{C}_{2y}\hat{\mathcal{T}}$ symmetries are derived from point group symmetries of $t$MoTe$_2$. On the other hand, $\hat{\mathcal{I}}$ is an emergent symmetry of the low-energy effective Hamiltonian $H$, and $\hat{\mathcal{I}}$ is a combination of $\hat{R}_{2z}$ and layer exchange $\sigma_x$.

The scalar potential $\Delta_0(\boldsymbol r)$ and the layer pseudospin  field $\boldsymbol\Delta(\boldsymbol r)$ are defined as
\begin{equation}
\label{Delta_de}
\begin{aligned}
\Delta_0(\boldsymbol r)= &\frac{\Delta_+(\boldsymbol r)+\Delta_-(\boldsymbol r)}{2},\\
\boldsymbol\Delta(\boldsymbol r)=  & [\text{Re}\,\widetilde\Delta_t^\dagger(\boldsymbol{r}),\text{Im}\,\widetilde\Delta_t^\dagger(\boldsymbol{r}),
\frac{\Delta_+(\boldsymbol r)-\Delta_-(\boldsymbol r)}{2}], \\
\widetilde\Delta_t^\dagger(\boldsymbol{r})=&e^{i(\boldsymbol \kappa_+-\boldsymbol   \kappa_-)\cdot \boldsymbol r}\Delta_t^\dagger(\boldsymbol r).
\end{aligned}
\end{equation}
$\boldsymbol\Delta(\boldsymbol r)$ transforms under $R_{3z}$, $R_{2z}$, $M_{x}$ and translational operations as follows,
\begin{equation}
\label{Delta_sym}
\begin{aligned}
&\boldsymbol\Delta(\hat{R}_{3z}\boldsymbol{r})=\boldsymbol\Delta(\boldsymbol{r}),\\
&\boldsymbol\Delta(\hat{R}_{2z}\boldsymbol{r})=[\Delta_x(\boldsymbol{r}),-\Delta_y(\boldsymbol{r}),-\Delta_z(\boldsymbol{r})],\\
&\boldsymbol\Delta(\hat{M}_x\boldsymbol{r})=[\Delta_x(\boldsymbol{r}),\Delta_y(\boldsymbol{r}),-\Delta_z(\boldsymbol{r})],\\
&\Delta_z(\boldsymbol{r}+\boldsymbol{a}_i)=\Delta_z(\boldsymbol{r}),\\
&\Delta_{x}(\boldsymbol{r}+\boldsymbol{a}_i)\pm i\Delta_{y}(\boldsymbol{r}+\boldsymbol{a}_i)=\\
&e^{\pm i(\boldsymbol{\kappa}_+-\boldsymbol{\kappa}_-)\cdot \boldsymbol{a}_i}[\Delta_{x}(\boldsymbol{r})\pm i\Delta_{y}(\boldsymbol{r})].
\end{aligned}
\end{equation}

We then consider the symmetries of $\alpha(\boldsymbol{r})$ and $\beta(\boldsymbol{r})$, which are defined as
 \begin{equation}
\begin{aligned}  
\alpha(\boldsymbol r)= & \frac{e^{-i  \zeta(\boldsymbol r)/2}}{\lvert\Psi^{(+)}_{0,\boldsymbol\kappa_-}(\boldsymbol r)\rvert}\sqrt{\frac{1+n_z(\boldsymbol r)}{2}}, \\
\beta(\boldsymbol r)= & \frac{e^{i\zeta(\boldsymbol r)/2}}{\lvert\Psi^{(+)}_{0,\boldsymbol\kappa_+}(\boldsymbol r)\rvert}
\sqrt{\frac{1-n_z(\boldsymbol r)}{2}}, \\
\zeta(\boldsymbol r)= & \text{Arg}\{\frac{n_x(\boldsymbol r)+in_y(\boldsymbol r)}{[\Psi^{(+)}_{0,\boldsymbol\kappa_-}(\boldsymbol r)]^*\Psi^{(+)}_{0,\boldsymbol\kappa_+}(\boldsymbol r)}\}. 
\end{aligned}
\end{equation}
Here $n_j(\bm r)=\Delta_j(\bm r)/|\bm \Delta(\bm r)|$.
Using Eqs.~\eqref{s0},~\eqref{s1} and~\eqref{Delta_sym}, we have
\begin{equation}
\begin{aligned}
&\zeta(\hat R_{3z}\boldsymbol{r})=\zeta(\boldsymbol{r}),\\
&\zeta(\hat{R}_{2z}\boldsymbol{r})=-\zeta(\boldsymbol{r})-\frac{4}{3}\pi,\\
&\zeta(\hat M_{x}\boldsymbol{r})=\zeta(\boldsymbol{r}),\\
&\zeta(\boldsymbol{r}+\boldsymbol{a}_i)=\zeta(\boldsymbol{r}),
\end{aligned}
\end{equation}
\begin{equation}
\label{ab_sym}
\begin{aligned} 
&\alpha(\hat R_{3z}\boldsymbol r)=\alpha(\boldsymbol r),\beta(\hat R_{3z}\boldsymbol r)=\beta(\boldsymbol r),\\
&\alpha(\hat{R}_{2z}\boldsymbol r)=e^{i\frac{2}{3}\pi}\beta(\boldsymbol r),\beta(\hat{R}_{2z}\boldsymbol{r})=e^{-i\frac{2}{3}\pi}\alpha(\boldsymbol{r}),\\
&\alpha(\hat M_{x}\boldsymbol r)=\beta^*(\boldsymbol r),\beta(\hat M_{x}\boldsymbol r)=\alpha^*(\boldsymbol r),\\
&\alpha(\boldsymbol r+\boldsymbol{a}_i)=\alpha(\boldsymbol r),\beta(\boldsymbol r+\boldsymbol{a}_i)=\beta(\boldsymbol r).
\end{aligned}
\end{equation}
We emphasize that $\alpha(\boldsymbol{r})$ and $\beta(\boldsymbol{r})$ are periodic functions with the moir\'e lattice periodicity. We plot the amplitude and phase of $\alpha(\boldsymbol{r})$ and $\beta(\boldsymbol{r})$ in Fig.~\ref{fig:8}, where the symmetry constraints in Eq.~\eqref{ab_sym} are obeyed.

The spinors $\boldsymbol{\chi}^{(s)}(\boldsymbol{r})$ are defined as
\begin{equation}
\begin{aligned}
\boldsymbol\chi^{(+)}(\boldsymbol r)=&
\{\alpha(\boldsymbol r)\Psi^{(+)}_{0,\boldsymbol \kappa_-}(\boldsymbol r),
\beta(\boldsymbol r)\Psi^{(+)}_{0,\boldsymbol \kappa_+}(\boldsymbol r)\}^T,\\
\boldsymbol\chi^{(-)}(\boldsymbol r)=&
\{\beta^*(\boldsymbol r)[\Psi^{(+)}_{0,\boldsymbol \kappa_+}(\boldsymbol r)]^*,
-\alpha^*(\boldsymbol r)[\Psi^{(+)}_{0,\boldsymbol \kappa_-}(\boldsymbol r)]^*\}^T.
\end{aligned}
\end{equation}
The transformation of $\boldsymbol{\chi}^{(s)}(\boldsymbol{r})$ under symmetry operations is given by, 
\begin{equation}
\label{chi_sym}
\begin{aligned}
&\boldsymbol{\chi}^{(s)}(\hat{R}_{3z}\boldsymbol{r})=\boldsymbol{\chi}^{(s)}(\boldsymbol{r}),\\
&\boldsymbol{\chi}^{(s)}
(\hat{R}_{2z}\boldsymbol{r})=-s\sigma_x\boldsymbol{\chi}^{(s)}(\boldsymbol{r}),\\
&\boldsymbol{\chi}^{(s)}(\hat{M}_{x}\boldsymbol{r})=-s\sigma_x[\boldsymbol{\chi}^{(s)}(\boldsymbol{r})]^*,\\
&\boldsymbol{\chi}^{(+)}(\boldsymbol{r}+\boldsymbol{a}_i)=-e^{-i\frac{1}{2\ell^2}\boldsymbol{a}_i\times\boldsymbol{r}}\widetilde U_0(\boldsymbol{a}_i)\boldsymbol{\chi}^{(+)}(\boldsymbol{r}),\\
&\boldsymbol{\chi}^{(-)}(\boldsymbol{r}+\boldsymbol{a}_i)=-e^{i\frac{1}{2\ell^2}\boldsymbol{a}_i\times\boldsymbol{r}}U_0^\dagger(\boldsymbol{a}_i)\boldsymbol{\chi}^{(-)}(\boldsymbol{r}),
\end{aligned}
\end{equation}
where $\widetilde U_0(\boldsymbol{r})=\mathrm{diag}(e^{i\boldsymbol{\kappa_-}\cdot \boldsymbol{r}},e^{i\boldsymbol{\kappa_+}\cdot \boldsymbol{r}})$. We note that $\boldsymbol{\chi}^{(s)}(\boldsymbol{r})$ obey magnetic translational symmetry.

The dressed Landau-level wavefunctions are given by
\begin{equation}
\begin{aligned}
\psi_{n,\boldsymbol  k}^{(+)}(\boldsymbol r)= & U_{0}(\boldsymbol r)\boldsymbol\chi^{(+)}(\boldsymbol r)\Psi_{n,\boldsymbol k}^{(-)}(\boldsymbol r), \\
\psi_{n,\boldsymbol  k}^{(-)}(\boldsymbol r)= & U_{0}(\boldsymbol r)\boldsymbol\chi^{(-)}(\boldsymbol r)\Psi_{n,\boldsymbol k}^{(+)}(\boldsymbol r).
\end{aligned}     
\end{equation}
Here $\Psi^{(s)}_{n,\boldsymbol k}(\boldsymbol r)$ is the wavefunction for the $n$th Landau level and satisfies the magnetic translational symmetry,
\begin{align}
\Psi_{n,\boldsymbol k}^{(s)}(\boldsymbol r+\boldsymbol{a}_i)
=-e^{-i\frac{1}{2\ell^2}s\boldsymbol {a}_i\times \boldsymbol r}e^{i\boldsymbol k\cdot \boldsymbol {a}_i}\Psi^{(s)}_{n,\boldsymbol k}(\boldsymbol r).
\end{align}
The combination of $U_0(\bm r)\boldsymbol{\chi}^{(s)}(\boldsymbol{r})$ and $\Psi_{0,\boldsymbol{k}}^{(-s)}(\boldsymbol{r})$ leads to normal translational symmetry, 
\begin{equation}
\begin{aligned}
\psi_{n,\boldsymbol{k}}^{(s)}(\boldsymbol{r}+\boldsymbol{a}_i)=&e^{i\boldsymbol{k}\cdot\boldsymbol{a}_i}\psi_{n,\boldsymbol{k}}^{(s)}(\boldsymbol{r}).
\end{aligned}
\end{equation}
Therefore, $\psi_{n,\boldsymbol{k}}^{(s)}(\boldsymbol{r})$ are Bloch wavefunctions.

Point group symmetries for the dressed zeroth Landau-level wavefunctions at the three high-symmetry momenta  are given by
\begin{equation}
\begin{aligned}
 \hat{C}_{3z}[U_0^\dagger(\boldsymbol{r})\psi^{(s)}_{0,\boldsymbol \gamma}(\boldsymbol r)]=&e^{i\frac{2}{3}\pi s}[U_0^\dagger(\boldsymbol{r})\psi^{(s)}_{0,\boldsymbol \gamma}(\boldsymbol r)],\\
 \hat{C}_{2y}\hat{\mathcal{T}}[U_0^\dagger(\boldsymbol{r})\psi^{(s)}_{0,\boldsymbol \gamma}(\boldsymbol r)]=&s[U_0^\dagger(\boldsymbol{r})\psi^{(s)}_{0,\boldsymbol \gamma}(\boldsymbol r)],\\
  \hat{\mathcal{I}}[U_0^\dagger(\boldsymbol{r})\psi^{(s)}_{0,\boldsymbol \gamma}(\boldsymbol r)]=&s [U_0^\dagger(\boldsymbol{r})\psi^{(s)}_{0,\boldsymbol \gamma}(\boldsymbol r)],\\
 \hat{C}_{3z}[U_0^\dagger(\boldsymbol{r})\psi^{(s)}_{0,\boldsymbol {\kappa}_\pm}(\boldsymbol r)]=&U_0^\dagger(\boldsymbol{r})\psi^{(s)}_{0,\boldsymbol {\kappa}_\pm}(\boldsymbol r),\\
 \hat{C}_{2y}\hat{\mathcal{T}}[U_0^\dagger(\boldsymbol{r})\psi^{(s)}_{0,\boldsymbol {\kappa}_\pm}(\boldsymbol r)]=&s[U_0^\dagger(\boldsymbol{r})\psi^{(s)}_{0,\boldsymbol {\kappa}_\mp}(\boldsymbol r)],\\
  \hat{\mathcal{I}}[U_0^\dagger(\boldsymbol{r})\psi^{(s)}_{0,\boldsymbol {\kappa}_\pm}(\boldsymbol r)]=&s e^{\mp i\frac{2}{3}\pi s}[U_0^\dagger(\boldsymbol{r})\psi^{(s)}_{0,\boldsymbol {\kappa}_\mp}(\boldsymbol r)].\\
\end{aligned}
\label{symmrepLL}
\end{equation}
Here $\hat{C}_{3z}=\hat{R}_{3z}$, $\hat{C}_{2y}\hat{\mathcal{T}}=\sigma_x \hat{M}_x \mathcal{K}$, $\hat{\mathcal{I}}=\sigma_x \hat{R}_{2z}$, and $\mathcal{K}$ is the complex conjugation operator.

\begin{figure}[t]
    \includegraphics[width=1.\columnwidth]{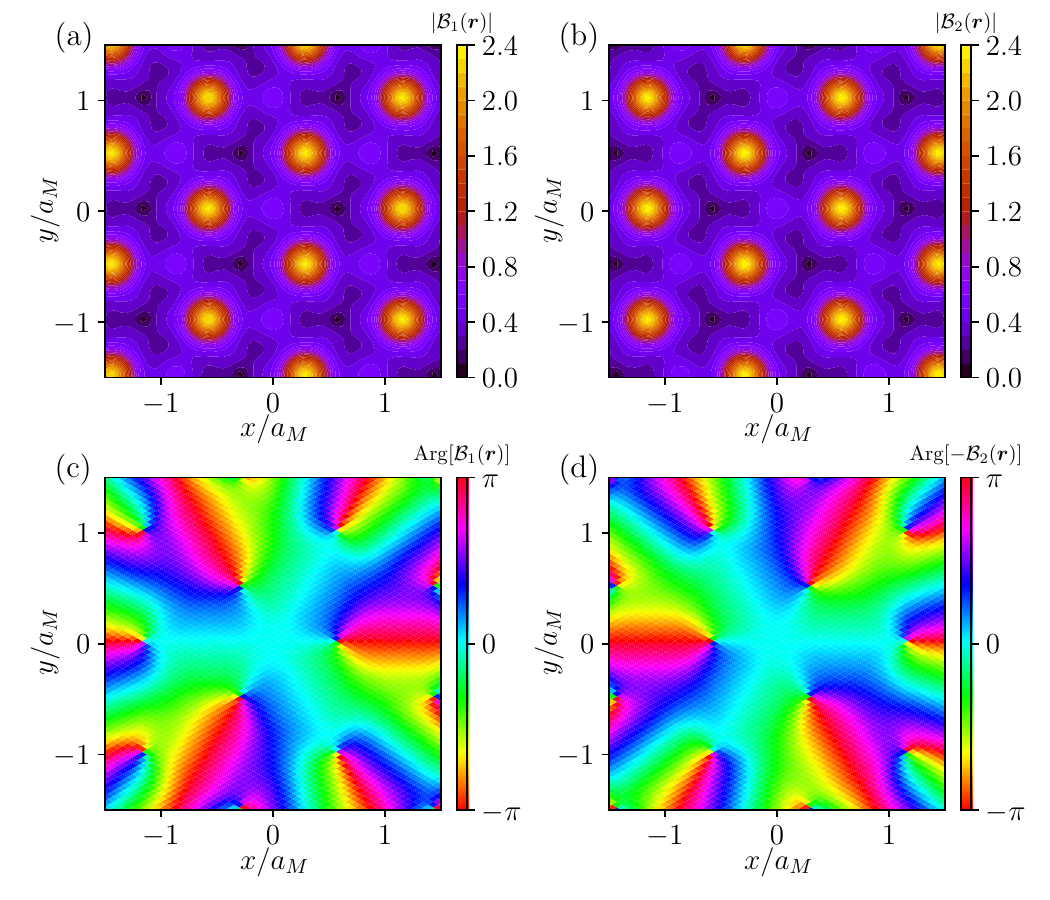}
    \caption{Spatial distribution of (a) $\lvert \mathcal{B}_1(\boldsymbol{r})\rvert$, (b) $\lvert \mathcal{B}_2(\boldsymbol{r})\rvert$, (c) $\mathrm{Arg}[\mathcal{B}_1(\boldsymbol{r})]$, and (d) $\mathrm{Arg}[-\mathcal{B}_2(\boldsymbol{r})]$ for $\theta=2.9^\circ$.}
    \label{fig:9}
\end{figure}

The original moir\'e Hamiltonian $H$ has the moir\'e periodicity. The point group symmetries are revealed more clearly using $H_1=U_{0}^\dagger(\boldsymbol r)HU_{0}(\boldsymbol r)$,
\begin{equation}
\label{H1_sym}
\begin{aligned}
& \hat C_{3z}H_1(\boldsymbol{r})\hat C_{3z}^{-1} \equiv H_1(\hat{R}_{3z}\boldsymbol{r})=H_1(\boldsymbol{r}),\\
& [\hat C_{2y}\hat {\mathcal{T}}]H_1(\boldsymbol{r})[\hat C_{2y}\hat{\mathcal{T}}]^{-1} \equiv \sigma_xH_1^*(\hat{M}_x\boldsymbol{r})\sigma_x=H_1(\boldsymbol{r})\,\\
& \hat{\mathcal{I}}H_1(\boldsymbol{r})\hat{\mathcal{I}}^{-1} \equiv \sigma_xH_1(\hat{R}_{2z}\boldsymbol{r})\sigma_x=H_1(\boldsymbol{r}).
\end{aligned}
\end{equation}

We find that the Bloch wavefunciton $\varphi_{1,\bm{k}}(\bm r)$ of the first moir\'e band of $H$ has the following symmetry representation at high-symmetry momenta,
\begin{equation}
\begin{aligned}
\hat{C}_{3z}[U_0^\dagger(\boldsymbol{r})\varphi_{1,\boldsymbol \gamma}(\boldsymbol r)]=&e^{i\frac{2}{3}\pi }[U_0^\dagger(\boldsymbol{r})\varphi_{1,\boldsymbol \gamma}(\boldsymbol r)],\\
  \hat{\mathcal{I}}[U_0^\dagger(\boldsymbol{r})\varphi_{1,\boldsymbol \gamma}(\boldsymbol r)]=&+ [U_0^\dagger(\boldsymbol{r})\varphi_{1,\boldsymbol \gamma}(\boldsymbol r)],\\
 \hat{C}_{3z}[U_0^\dagger(\boldsymbol{r})\varphi_{1,\boldsymbol {\kappa}_\pm}(\boldsymbol r)]=&+[U_0^\dagger(\boldsymbol{r})\varphi_{1,\boldsymbol {\kappa}_\pm}(\boldsymbol r)].
\end{aligned}
\label{symmrep}
\end{equation}
By comparing Eqs.~\eqref{symmrepLL} and \eqref{symmrep}, we conclude that the dressed zeroth Landau-level wavefunction $\psi_{0,\bm k}^{(+)}(\bm r)$ consistently captures the symmetry representation of the first moir\'e band wavefunction $\varphi_{1,\bm{k}}(\bm r)$.

Under lattice translation, $H_1$ transforms as follows,
\begin{equation}
\begin{aligned}
H_1(\boldsymbol{r}+\boldsymbol{a}_i)=U_0^\dagger(\boldsymbol{a}_i)H_1(\boldsymbol{r})U_0(\boldsymbol{a}_i).
\end{aligned}
\end{equation}
In the local frame of $\boldsymbol{\Delta}(\boldsymbol{r})$, the Hamiltonian is transformed to $H_2=U^\dagger (\boldsymbol{r})H_1U(\boldsymbol{r})$, where $U(\boldsymbol r)=  (\boldsymbol\chi^{(+)}(\boldsymbol r),\boldsymbol\chi^{(-)}(\boldsymbol r))$. The transfomation of $U(\boldsymbol r)$ under lattice translation is,
\begin{equation}
\begin{aligned}
U(\boldsymbol{r}+\boldsymbol{a}_i)=-U_0^\dagger(\boldsymbol{a}_i)U(\boldsymbol{r})U_m(\boldsymbol{r})
\end{aligned}
\end{equation}
where $U_m(\boldsymbol{r})$ is
\begin{equation}
U_m(\boldsymbol{r})=
\begin{pmatrix}
e^{-i\frac{1}{2\ell^2}\boldsymbol{a}_i\times \boldsymbol{r}} & 0\\
0 &  e^{i\frac{1}{2\ell^2}\boldsymbol{a}_i\times \boldsymbol{r}}.
\end{pmatrix}
\end{equation}
Therefore, $H_2$ has the following translational symmetry,
\begin{equation}
\begin{aligned}
H_2(\boldsymbol{r}+\boldsymbol{a}_i)=U_m^\dagger(\boldsymbol{r})H_2(\boldsymbol{r})U_m(\boldsymbol{r}).
\end{aligned}
\end{equation}

We plot in Fig.~\ref{fig:9} the spatial distribution of $\lvert \mathcal{B}_1(\boldsymbol{r})\rvert$, $\lvert \mathcal{B}_2(\boldsymbol{r})\rvert$, $\mathrm{Arg}[\mathcal{B}_1(\boldsymbol{r})]$ and $\mathrm{Arg}[-\mathcal{B}_2(\boldsymbol{r})]$ for $\theta=2.9^\circ$. We note that $\mathcal{B}(\boldsymbol{r})$ inherits the symmetries of $\boldsymbol{\chi}^{(+)}(\boldsymbol{r})$.

\section{Variational method}
\label{appendix:B}
The details of the variational method are described as follows. The expression of $\widetilde c_{\boldsymbol{k}}$ is given by
\begin{equation}
\begin{aligned}
 \widetilde c_{\boldsymbol{k}} &=  \langle\varphi_{1,\boldsymbol k}\lvert\Theta_{\boldsymbol k}\rangle\\ 
&=  \frac{\sum_{i=1,2}\int_{\mathcal{A}_0} d\boldsymbol{r}\,\varphi^*_{1,\boldsymbol{k},i}(\boldsymbol{r})[U_0(\boldsymbol{r})]_{ii}\mathcal{B}_i(\boldsymbol{r})\Psi_{0,\boldsymbol{k}}^{(-)}(\boldsymbol{r})}{\sqrt{\int_{\mathcal{A}_0} d\boldsymbol{r}\,\lvert \mathcal{B}(\boldsymbol{r})\rvert^2\lvert\Psi_{0,\boldsymbol{k}}^{(-)}(\boldsymbol{r})\rvert^2}\sqrt{\int_{\mathcal{A}_0} d\boldsymbol{r}\,\lvert\varphi_{1,\boldsymbol{k}}(\boldsymbol{r})\rvert^2}}\\
& = \frac{\sum_{i=1,2}\sum_{\boldsymbol{r}}\varphi^*_{1,\boldsymbol{k},i}(\boldsymbol{r})[U_0(\boldsymbol{r})]_{ii}\mathcal{B}_i(\boldsymbol{r})\Psi_{0,\boldsymbol{k}}^{(-)}(\boldsymbol{r})}{\sqrt{\sum_{\boldsymbol{r}}\lvert \mathcal{B}(\boldsymbol{r})\rvert^2\lvert\Psi_{0,\boldsymbol{k}}^{(-)}(\boldsymbol{r})\rvert^2}}
\end{aligned}   
\end{equation}
where we take the normalization $\sum_{\boldsymbol{r}}\lvert\varphi_{1,\boldsymbol{k}}(\boldsymbol{r})\rvert^2=1$, with discretized summation within a unit cell. The weight $\widetilde W$ takes the form 
\begin{equation}
\begin{aligned}
\widetilde W = & \frac{1}{N}\sum_{\boldsymbol{k}} \lvert \widetilde c_{\boldsymbol{k}} \rvert^2 = \frac{1}{N}\sum_{\boldsymbol{k}}\frac{\lvert u_{\boldsymbol{k}}\rvert^2}{v_{\boldsymbol{k}}},\\
u_{\boldsymbol{k}}=&\sum_{i=1,2}\sum_{\boldsymbol{r}}\varphi^*_{1,\boldsymbol{k},i}(\boldsymbol{r})[U_0(\boldsymbol{r})]_{ii}\mathcal{B}_i(\boldsymbol{r})\Psi_{0,\boldsymbol{k}}^{(-)}(\boldsymbol{r}),\\
v_{\boldsymbol{k}}=&\sum_{\boldsymbol{r}}\lvert \mathcal{B}(\boldsymbol{r})\rvert^2\lvert\Psi_{0,\boldsymbol{k}}^{(-)}(\boldsymbol{r})\rvert^2.
\end{aligned}    
\end{equation}
The gradients of $u_{\boldsymbol{k}}$ and $v_{\boldsymbol{k}}$ with respect to $\mathcal{B}_i(\boldsymbol{r})$  are
\begin{equation}
\begin{aligned}
\frac{\delta u_{\boldsymbol{k}}}{\delta \mathrm{Re}[\mathcal{B}_i(\boldsymbol{r})]}= & \varphi^*_{1,\boldsymbol{k},i}[U_{0}(\boldsymbol{r})]_{ii}(\boldsymbol{r})\Psi_{0,\boldsymbol{k}}^{(-)}(\boldsymbol{r}), \\
\frac{\delta u_{\boldsymbol{k}}}{\delta \mathrm{Im}[\mathcal{B}_i(\boldsymbol{r})]}= & i\varphi^*_{1,\boldsymbol{k},i}(\boldsymbol{r})[U_{0}(\boldsymbol{r})]_{ii}\Psi_{0,\boldsymbol{k}}^{(-)}(\boldsymbol{r}),  \\
\frac{\delta v_{\boldsymbol{k}}}{\delta \mathrm{Re}[\mathcal{B}_i(\boldsymbol{r})]}= & 2\mathrm{Re}[\mathcal{B}_i(\boldsymbol{r})]\lvert\Psi_{0,\boldsymbol{k}}^{(-)}(\boldsymbol{r})\rvert^2, \\
\frac{\delta v_{\boldsymbol{k}}}{\delta \mathrm{Im}[\mathcal{B}_i(\boldsymbol{r})]}= & 2\mathrm{Im}[\mathcal{B}_i(\boldsymbol{r})]\lvert\Psi_{0,\boldsymbol{k}}^{(-)}(\boldsymbol{r})\rvert^2. 
\end{aligned}
\end{equation}
The gradients of $\widetilde W$ are then given by,
\begin{equation}
\label{dev1}
\begin{small}
\begin{aligned} 
&\frac{\delta\widetilde W}{\delta \mathrm{Re}[\mathcal{B}_i(\boldsymbol{r})]}=\frac{1}{N}\sum_{\boldsymbol{k}}\frac{1}{v_{\boldsymbol{k}}^2}[\frac{\delta \lvert u_{\boldsymbol{k}}\rvert^2}{\delta \mathrm{Re}[\mathcal{B}_i(\boldsymbol{r})]}v_{\boldsymbol{k}}-\lvert u_{\boldsymbol{k}}\rvert^2\frac{\delta v_{\boldsymbol{k}}}{\delta \mathrm{Re}[\mathcal{B}_i(\boldsymbol{r})]}], \\
&\frac{\delta\widetilde W}{\delta \mathrm{Im}[\mathcal{B}_i(\boldsymbol{r})]}= \frac{1}{N}\sum_{\boldsymbol{k}}\frac{1}{v_{\boldsymbol{k}}^2}[\frac{\delta \lvert u_{\boldsymbol{k}}\rvert^2}{\delta \mathrm{Im}[\mathcal{B}_i(\boldsymbol{r})]}v_{\boldsymbol{k}}-\lvert u_{\boldsymbol{k}}\rvert^2\frac{\delta v_{\boldsymbol{k}}}{\delta \mathrm{Im}[\mathcal{B}_i(\boldsymbol{r})]}].
\end{aligned}
\end{small}
\end{equation}

In our variational method, we set the parameter $\xi=0.5$ and iterate until the difference in $\widetilde W$ between two consecutive steps is less than $10^{-8}$.

\bibliography{ref}

\end{document}